\documentclass[acmtog,nonacm]{acmart}
\acmSubmissionID{996}

\AtBeginDocument{%
  }

\setcopyright{acmlicensed}
\copyrightyear{2026}
\acmYear{2026}

\acmConference[SIGGRAPH 2026]{}{}{}

\usepackage{graphicx}
\usepackage{array}
\usepackage{booktabs}
\usepackage{makecell}

\begin{document}

\title{StyleID: A Perception-Aware Dataset and Metric for Stylization-Agnostic Facial Identity Recognition}

\author{Kwan Yun}
\affiliation{%
  \institution{KAIST}
  \country{South Korea}
}
\email{yunandy@kaist.ac.kr}
\author{Changmin Lee}
\affiliation{%
  \institution{KAIST}
  \country{South Korea}
}
\author{Ayeong Jeong}
\affiliation{%
  \institution{KAIST}
  \country{South Korea}
}
\author{Youngseo Kim}
\affiliation{%
  \institution{KAIST}
  \country{South Korea}
}
\author{Seungmi Lee}
\affiliation{%
  \institution{KAIST}
  \country{South Korea}
}
\author{Junyong Noh} 
\affiliation{%
  \institution{KAIST}
  \country{South Korea}
}
\renewcommand\shortauthors{Yun et al.}

\ccsdesc[500]{Computing methodologies~Perception}
\begin{teaserfigure}
  \includegraphics[width=\textwidth]{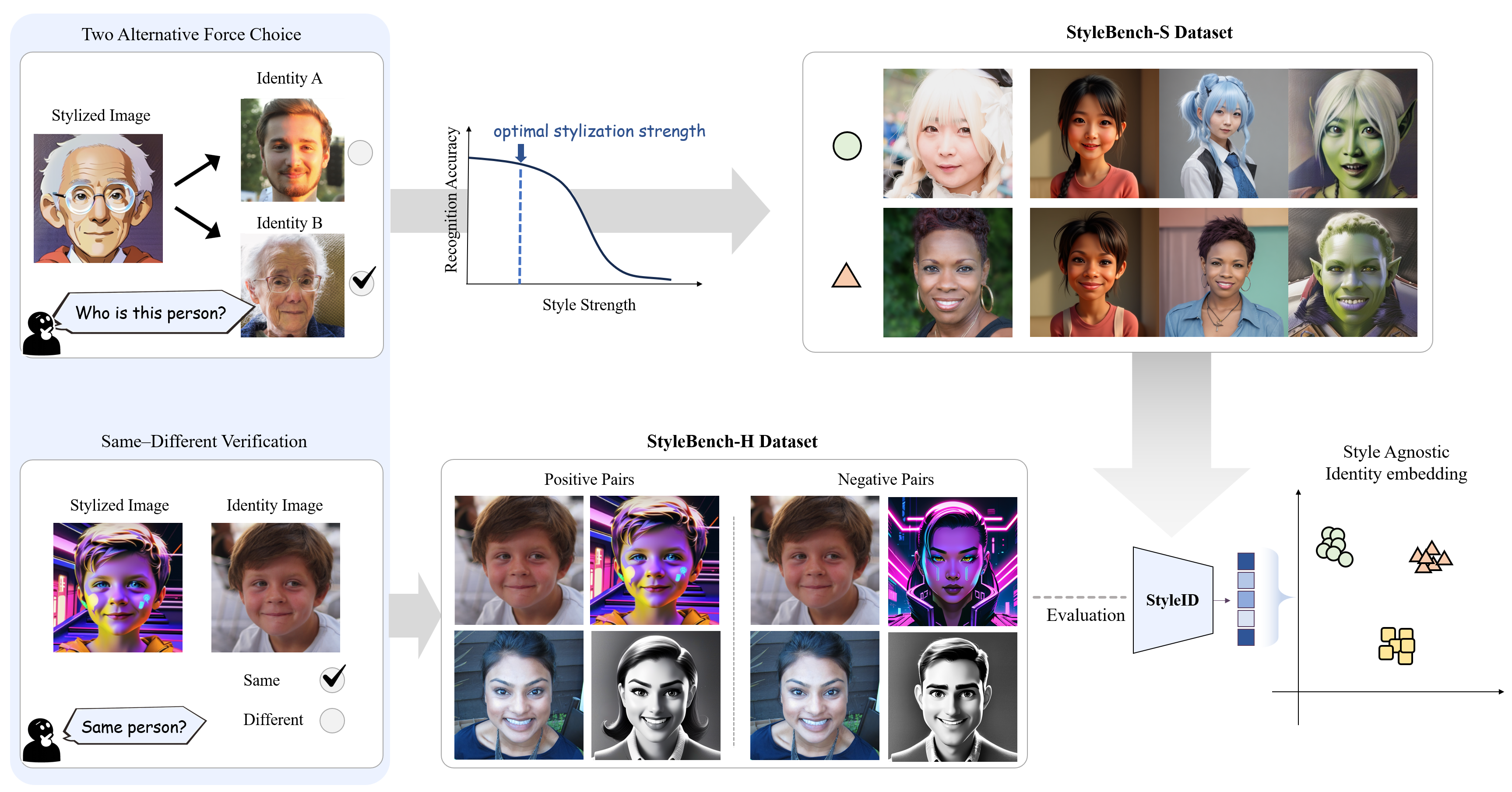}\vspace{-2mm}
  \caption{\textbf{StyleBench-H} is a human-judged benchmark for identity verification under controllable face stylization, and \textbf{StyleBench-S} is a large-scale synthetic supervision set derived from human-calibrated recognition statistics. \textbf{StyleID} is trained on StyleBench-S, enabling robust identity recognition across diverse styles while remaining aligned with human judgment. The resulting model supports identity recognition, stylization, and retrieval tasks.
  }
  \Description{}
  \label{fig:teaser}
\end{teaserfigure}


\newcommand{\fixq}[1]{\textcolor{black}{#1}}
\newcommand{\fixw}[1]{\textcolor{black}{#1}}

\begin{abstract}
Creative face stylization aims to render portraits in diverse visual idioms such as cartoons, sketches, and paintings while retaining recognizable identity. However, current identity encoders, which are typically trained and calibrated on natural photographs, exhibit severe brittleness under stylization. They often mistake changes in texture or color palette for identity drift or fail to detect geometric exaggerations. This reveals the lack of a style-agnostic framework to evaluate and supervise identity consistency across varying styles and strengths. To address this gap, we introduce StyleID, a human perception-aware dataset and evaluation framework for facial identity under stylization. StyleID comprises two datasets: (i) StyleBench-H, a benchmark that captures human same–different verification judgments across diffusion- and flow-matching-based stylization at multiple style strengths, and (ii) StyleBench-S, a supervision set derived from psychometric recognition–strength curves obtained through controlled two-alternative forced-choice (2AFC) experiments. Leveraging StyleBench-S, we fine-tune existing semantic encoders to align their similarity orderings with human perception across styles and strengths. Experiments demonstrate that our calibrated models yield significantly higher correlation with human judgments and enhanced robustness for out-of-domain, artist drawn portraits. All of our datasets, code, and pretrained models are publicly available at \href{https://kwanyun.github.io/StyleID_page/}{\textbf{StyleID}}.
\end{abstract}

\maketitle

\section{Introduction}
\label{sec:intro}
Creative face stylization, which can be defined as rendering a portrait in the aesthetic of a cartoon, painting, sketch, or other visual idiom, has become routine in consumer apps, avatar platforms, professional content pipelines, and \fixq{multimodal} language models such as ChatGPT~\cite{chatgpt} and Gemini~\cite{gemini}. The aspect of identity preservation is typically important in these stylizations because recognizability under style transfer is central to user identity, personalization, and social presence. Therefore, personalized avatars require identity metrics that remain reliable under distribution shift induced by style. 

However, most existing identity encoders and their verification thresholds are trained and calibrated on natural photographs and evaluated on in-distribution photo benchmarks. Despite this narrow training distribution, these models are routinely repurposed both as evaluators of stylized face and as identity-preservation objectives within learning-based stylization pipelines. In practice, their scores can be brittle under stylization that they may mistake changes in texture or color palette for identity change or, conversely, fail to detect identity drift introduced by exaggerated geometry.

We attribute this brittleness to a missing piece: a style-agnostic, human-calibrated protocol for measuring identity consistency across stylization methods and across stylization strengths. This gap has persisted even as face stylization has rapidly evolved—from pre-trained CNN style transfer~\cite{gatys2016image,johnson2016perceptual,ulyanov2016instance}, to GAN-based image translation~\cite{zhu2017unpaired,isola2017image}, and more recently to diffusion- and flow-matching–based editing~\cite{ye2023ip,wang2024instantid,brooks2023instructpix2pix,wu2025qwen,batifol2025flux}. Current evaluations typically either (i) reuse photo-domain encoders and thresholds that do not reliably transfer to stylized images~\cite{deng2019arcface,kim2022adaface}, or (ii) rely on generic similarity measures that are not designed to match human-perceived identity (e.g., CLIP)~\cite{radford2021clip}. While StylizedFace~\cite{peng2025stylized} recently proposed an evaluation approach, it is neither calibrated to human judgment nor publicly available, limiting its practical utility. As a result, developers lack an evidence-based means to compare stylization pipelines in terms of identity preservation, and researchers lack supervision that anchors identity metrics to human judgments under stylization.

We address this gap with StyleID: a perception-aware dataset and evaluation framework for identity under face stylization. As illustrated in Figure~\ref{fig:teaser}, our key idea is to place human perception at the center of the evaluation and to explicitly factor in stylization by method and strength. Concretely, we consider three recent diffusion- and flow-matching–based stylization methods~\cite{ye2023ip,wang2024instantid,jiang2025infiniteyou}. Each induces a controllable style strength parameter that modulates deviation from the original portrait. For the evaluation of existing metrics whether it matches with human perception, we build a dataset StyleBench-H by stylizing the source human images using each family at multiple strengths and collect human identity judgments through same–different verification. Using StyleBench-H, we demonstrate existing metrics, widely used for identity preservation for stylized faces do not match with human perception.

Beyond evaluation, to build a metric that better matches human perception, we augment our benchmark with supervision by collecting pairwise similarity responses under controlled two-alternative forced-choice (2AFC) protocols. These responses yield recognition–strength curves—human recognition accuracy as a function of style strength for each stylization family—which we treat as psychometric functions describing how identity becomes ambiguous as stylization intensifies. Leveraging these curves, we derive a style-conditional calibration that maps a given stylization setting to a probability of “same identity”. We then generate synthetic training pairs that humans would likely judge as the same identity, along with hard negatives sampled near the ambiguity boundary. We denote this supervision set as StyleBench-S.

Using StyleBench-S, we fine-tune a semantic encoder CLIP~\cite{radford2021clip} with LoRA~\cite{hu2022lora}, combining a contrastive training strategy with angular identity losses so that similarity orderings align with human judgments across styles and strengths. Empirically, the calibrated and CLIP-adapted model, named StyleID, exhibits substantially higher correlation with human judgments and generalizes better to stylized portraits drawn by artists. In addition, we demonstrate that StyleID consistently outperforms prior identity encoders in recognition and retrieval tasks, and serves as a stronger supervisory signal for learning-based stylization models, which conventionally rely on identity encoders trained only on real human faces.

\section{Related Work}
\begin{figure*}
  \includegraphics[width=0.95\linewidth]{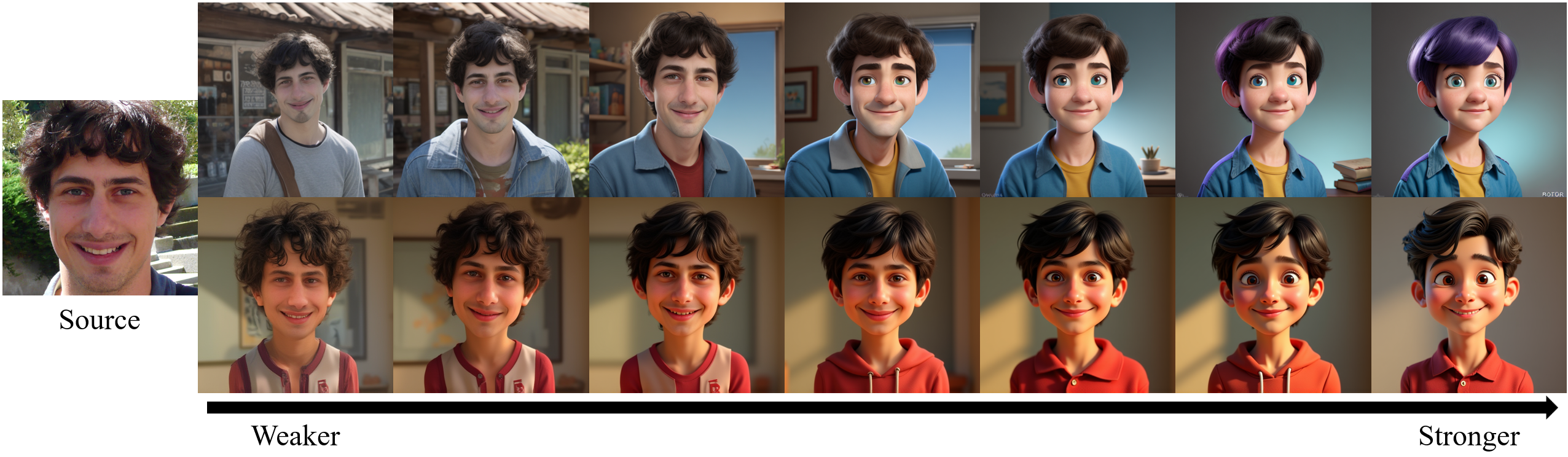}\vspace{-3mm}
  \caption{Example data generated for stylization from weaker to stronger stylization strengths. As the stylization strength increases, identity preservation decreases. The first row shows results from IP-Adapter, and the second row shows results from InfiniteYou.}
  \label{fig:example_data}
\end{figure*}

\subsection{Identity Recognition Model}
Identity recognition models aim to embed facial images into a feature space where distances reflect identity similarity—faces of the same person cluster closely while those of different people remain well separated. Early deep-learning approaches framed this as a classification problem with softmax losses~\cite{taigman2014deepface,parkhi2015deep}. To enhance discriminability, subsequent methods introduced constraints that directly shape the geometry of the embedding space. The center loss~\cite{wen2016discriminative} reduced intra-class variance by pulling features toward learned class centers, while triplet loss~\cite{schroff2015facenet} optimized relative distances among triplets of samples.

Modern metrics reformulate identity learning as margin-based classification on a normalized hypersphere. SphereFace~\cite{liu2017sphereface} introduced a multiplicative angular margin, later simplified into additive forms such as CosFace~\cite{wang2018cosface} and ArcFace~\cite{deng2019arcface}. These additive-margin softmax objectives enforce clear angular decision boundaries and became the de-facto standard for face recognition benchmarks. MagFace~\cite{meng2021magface} further extended this framework by coupling the embedding magnitude with face-image quality, allowing the network to learn quality-adaptive margins and to estimate sample reliability. Recent works build on these principles to address data imbalance, uncertainty, or robustness. ElasticFace~\cite{boutros2022elasticface} introduces stochastic margins to improve generalization, and AdaFace~\cite{kim2022adaface} adapts the margin dynamically based on estimated image quality. Despite these advances, most existing identity encoders are trained and calibrated on large photo-domain datasets, assuming naturalistic textures and photometric cues. As a result, these metrics exhibit unstable similarity scores and poor calibration when applied to stylized portraits. Most closely related to our work, StylizedFace~\cite{peng2025stylized} proposes generating stylized face images and training a recognition model on them. However, it relies on a single stylization strength and lacks calibration with human perception, limiting its alignment with human judgment. Our work builds upon this lineage but repositions identity measurement within a perceptual and style-aware framework by building human perception centric datasets and fine-tuning image encoders on them.

\subsection{Face Recognition Dataset}
Modern identity encoders are trained on large photo-domain face recognition \fixq{datasets} such as
CASIA-WebFace, MS-Celeb-1M/MS1MV2, and VGGFace2, as well as newer web-scale datasets such as WebFace42M and Glint360K, which enable margin-based training at scale~\cite{huang2008labeled,yi2014learning,guo2016ms,cao2018vggface2,deng2019arcface,zhu2021webface260m,an2021partial}. Despite their breadth, these resources remain photographic, and identity metrics calibrated on them can be brittle under stylization. While FaRL~\cite{zheng2022general} constructed LAION-FACE and leveraged image-text representations, its scope is still limited to natural faces. Our work complements this literature with perception-centered data and style-conditional protocols tailored to stylized portraits.

\subsection{Face Stylization}
Unlike classical style transfer approaches that match feature statistics to impose texture patterns~\cite{gatys2016image,johnson2016perceptual,ulyanov2016instance}, face stylization typically relies on learned domain mappings. Early portrait stylization and cartoonization works often relied on general-purpose image-to-image translation frameworks~\cite{isola2017pix2pix,zhu2017unpaired}, while later methods leverage GAN priors for stylization~\cite{yun2024stylized,men2022dct,jang2024toonify3d,kim2022dynagan,karras2020styleganada,tov2021e4e,roich2022pivotal} or incorporate geometry-aware modules to warp facial shape~\cite{shi2019warpgan,jang2021stylecarigan}.
Recent diffusion and flow/consistency models enable promptable and controllable stylization of real faces through inversion and editing pipelines~\cite{meng2021sdedit,hertz2022prompt,mokady2023null} as well as instruction- or exemplar-based conditioning~\cite{brooks2023instructpix2pix,zhang2023adding,ye2023ip}. Flow-matching and consistency generators further enhance editability under strong style transformations~\cite{wu2025qwen,batifol2025flux,flux-2-2025}.
Identity preservation is typically enforced through face-ID losses~\cite{deng2019arcface} or latent-space constraints during inversion; however, because these encoders are trained on real photos, their reliability degrades as stylization intensity increases or geometry is heavily exaggerated—highlighting the need for our perception-centered, style-aware evaluation.

\section{Dataset Construction} \label{sec:dataset}
To bridge the gap between algorithmic identity metrics and human perception, we construct StyleBench, a dataset comprising two distinct subsets: \textbf{StyleBench-H} (Human), a rigorous benchmark of human identity judgments across varying style strengths, and \textbf{StyleBench-S} (Synthetic), a large-scale synthetic dataset for model calibration and fine-tuning.

\subsection{Controllable Stylization Pipeline} \label{subsec:pipeline}
A core requirement of our analysis is the ability to generate stylized portraits with a controllable degree of deviation from the source identity. We employ three state-of-the-art diffusion-based and flow-based stylization frameworks. Specifically, we adopt IP-Adapter~\cite{ye2023ip}, InstantID~\cite{wang2024instantid}, and InfiniteYou~\cite{jiang2025infiniteyou}, which are widely used and support explicit control over stylization strength.

Firstly, IP-Adapter~\cite{ye2023ip} decouples cross-attention mechanisms to inject identity information into a pretrained text-to-image diffusion model, allowing stylization. We use \fixq{IP-Adapter-faceID,} a variant that injects the source identity and text prompt as conditioning signals. Stylization strength is controlled by the scale of the added attention layers and the text conditioning, denoted as $s_{ip} \in [0,1]$.
Secondly, InstantID combines ControlNet~\cite{zhang2023adding} with an IP-Adapter-like attention-injection mechanism. Here, stylization strength is controlled through the ControlNet strength and the style conditioning, which we denote as $s_{id} \in [0,1]$.
Lastly, InfiniteYou~\cite{jiang2025infiniteyou} is a recent flow-matching approach that enables high-fidelity generation with efficient sampling. We denote its style-strength parameter as $s_{inf} \in [0,1]$.

For all three methods, we normalize the strength parameters such that $s=0$ corresponds to marginal stylization that closely approximates the original source identity, while $s=1$ corresponds to maximum stylization with potential identity loss. For efficient dataset construction, we discretize the strength into seven levels for all methods. Because the mechanisms controlling style strength differ across methods, the same normalized value does not necessarily yield comparable perceptual strength (e.g., $s_{ip}=0.5$ does not imply perceptual equivalence to $s_{inf}=0.5$). Example outputs at varying strengths are shown in Figure~\ref{fig:example_data}.

\begin{figure}[t]
\hspace{-2mm}
  \includegraphics[width=1.03\linewidth]{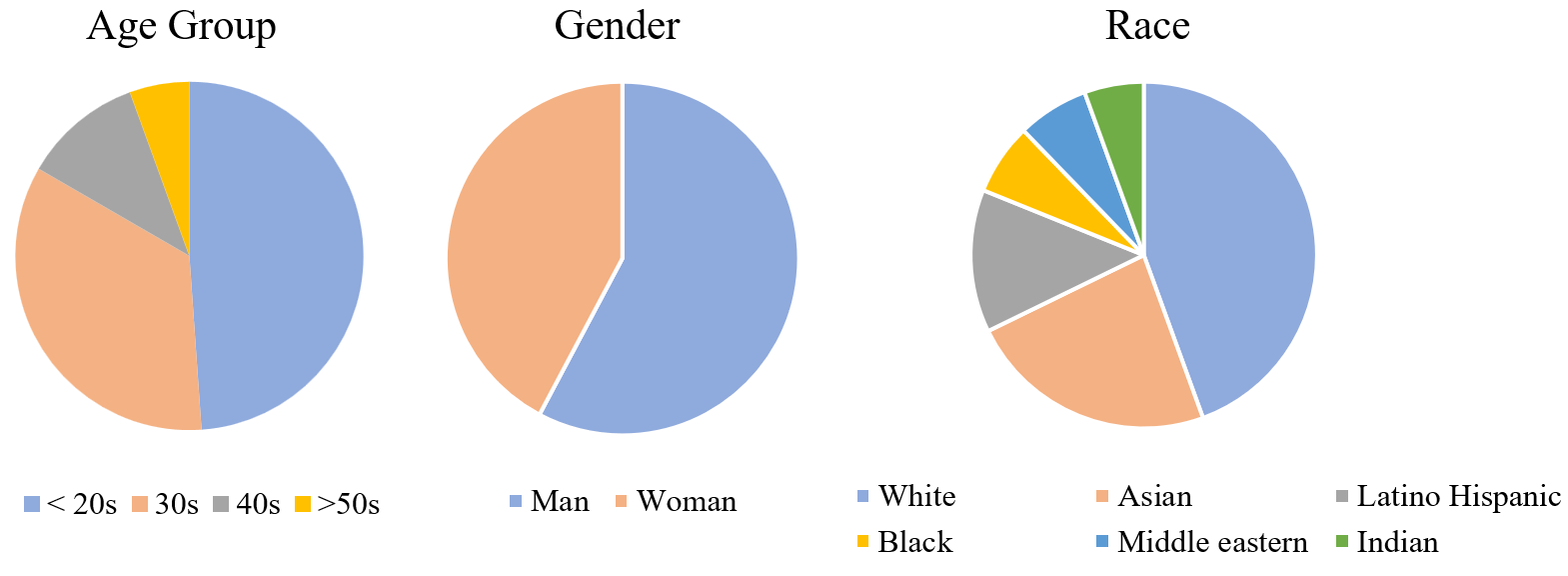}\vspace{-2mm}
  \caption{Demographics of StyleBench-H source images.}
  \label{fig:demographics}
\end{figure}

\subsection{StyleBench-H: Human Perception Benchmark} 
To evaluate how well face-recognition model scores align with human perception, we curate \textbf{StyleBench-H}. We sample high-quality source portraits with diverse racial and age attributes from FFHQ~\cite{karras2019stylegan}, filtering out images with large head rotations and those containing more than one person. We then apply the three stylization methods described above to the remaining source images. The demographics of the StyleBench-H source images are presented in Figure~\ref{fig:demographics}, which are extracted using a pretrained facial attribute analysis network~\cite{serengil2021lightface}.

For each source identity, we generate stylized counterparts using each stylization method across 10 artistic styles $t$ (e.g., watercolor, cyberpunk, anime) and seven discrete strength levels $s \in \{1/7, 2/7, \ldots, 7/7\}$. This design makes the benchmark robust to variation in style and texture. The resulting pairs $(I_{src}, I_{style})$ are presented to human annotators in a pairwise verification task. Annotators answer the question, “Do these two images depict the same person?”, with the explicit instruction that the second image is a stylistic rendering.

\begin{figure}[t]
\centering
  \includegraphics[width=1.02\linewidth]{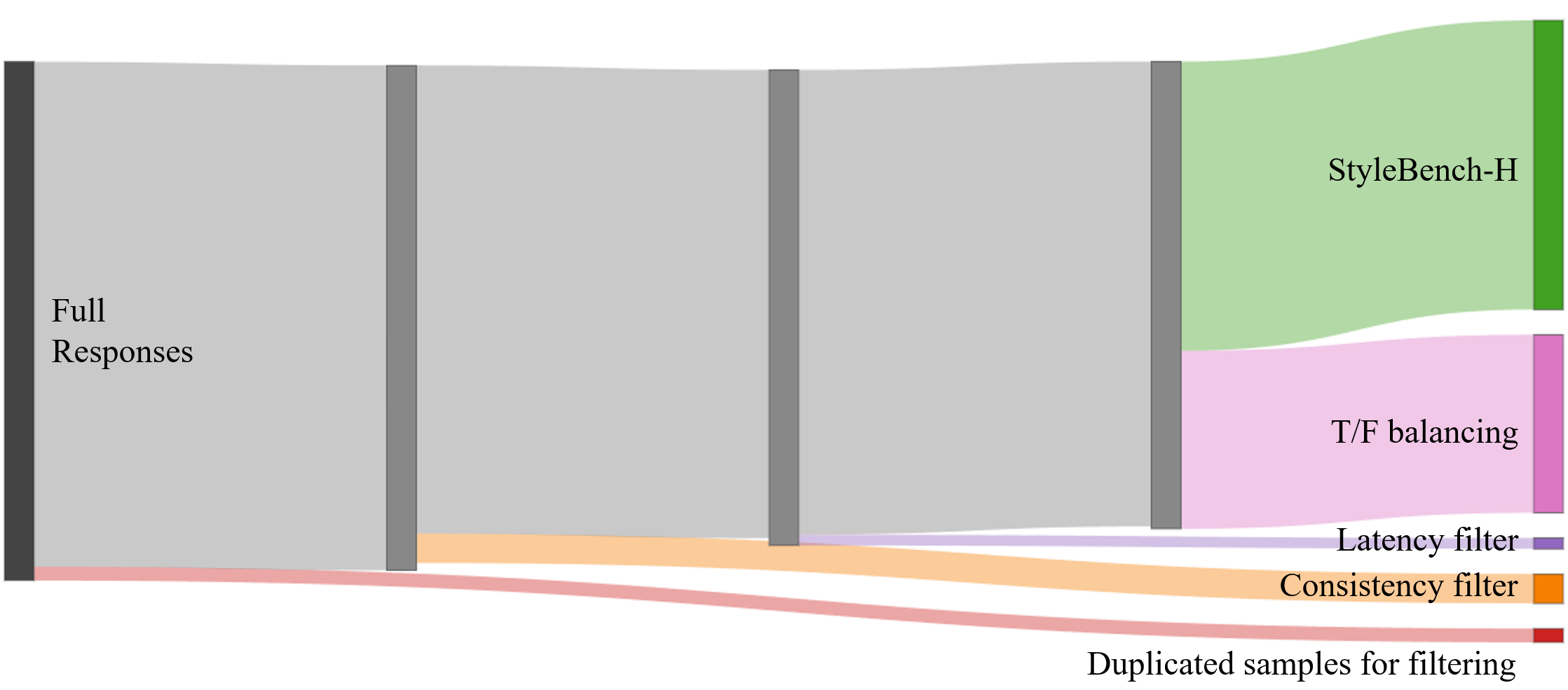}
  \caption{StyleBench-H dataset filtering pipeline.}
  \label{fig:styleid_filter}
\end{figure}

We recruited $N=70$ participants (Male: 32, Female: 33, Not disclosed: 5) with a mean age of 29.0 years, and each participant answered 91 queries. To ensure data quality, we applied filters based on response latency and consistency. For latency, we discarded responses submitted faster than the image loading time or those that took longer than 100 seconds for a single question. To filter out random guessing, we removed all data from participants who provided inconsistent answers to repeated questions. The repeated item was presented as the final question and was identical to the penultimate question, with no changes to the samples or their order. Responses from two participants were discarded, leaving a total of 68 valid participants. The stimuli were presented in a balanced distribution across the three methods and strength levels. From 6088 valid responses, we retained only true-positive and true-negative pairs and further balanced positives and negatives, yielding a total of $N_H=3551$ valid datapoints. The data retention and exclusion statistics are visualized as a Sankey diagram in Figure~\ref{fig:styleid_filter}.

While StyleBench-H primarily evaluates generalization to unseen identities, the stylization styles and methods are not fully disjoint from those used in StyleBench-S training. To further assess robustness under stricter distribution shifts, we additionally construct \fixq{Cross-Style} and \fixq{Cross-Method} splits for the full StyleBench-H. For these splits, we recruited $N=28$ participants (Male: 15, Female: 13) with a mean age of 30.4 years. In the \fixq{Cross-Style} setting, we use IP-Adapter to generate eight distinct styles for identities that do not overlap with those seen during training. In the \fixq{Cross-Method} setting, the test identities, stylization methods, and their associated prompts/styles are all unseen. Concretely, we introduce two unseen stylization methods, MTG~\cite{zhu2021mind} and Flux.2~\cite{flux-2-2025}, each paired with eight unseen styles, following a data collection and filtering procedure consistent with StyleBench-H \fixq{cross-ID}. These additional splits enable a more rigorous evaluation of style-agnostic identity consistency under out-of-domain stylization conditions.

\begin{figure*}[t]
\vspace{-2mm}
    \centering
    \begin{minipage}[t]{0.31\linewidth}
        \centering
        \includegraphics[width=\linewidth]{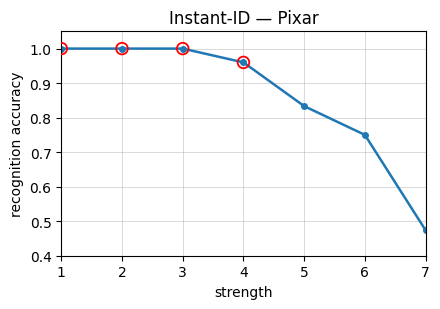}
    \end{minipage}
    \hfill
    \begin{minipage}[t]{0.31\linewidth}
        \centering
        \includegraphics[width=\linewidth]{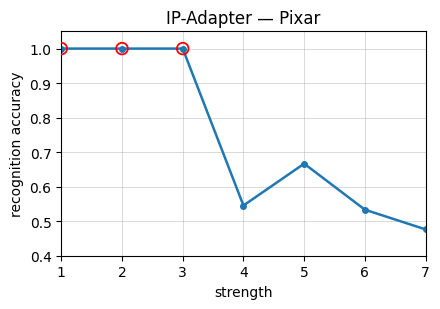}
    \end{minipage}
    \hfill
    \begin{minipage}[t]{0.31\linewidth}
        \centering
        \includegraphics[width=\linewidth]{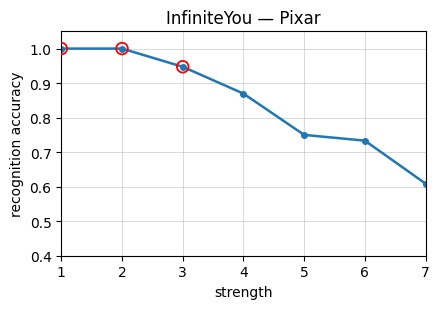}
    \end{minipage}

    \vspace{0.6em}

    \begin{minipage}[t]{0.31\linewidth}
        \centering
        \includegraphics[width=\linewidth]{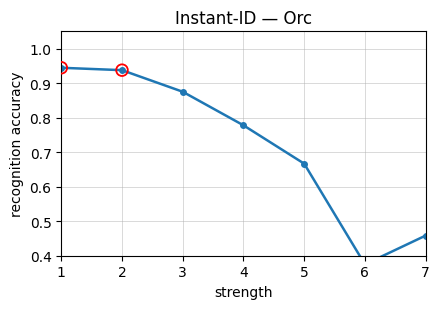}
    \end{minipage}
    \hfill
    \begin{minipage}[t]{0.31\linewidth}
        \centering
        \includegraphics[width=\linewidth]{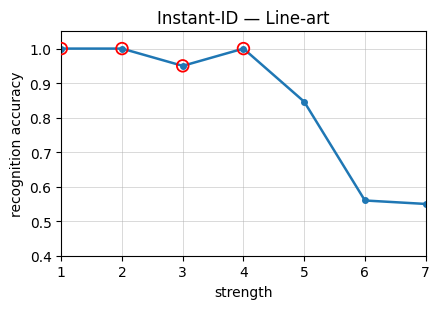}
    \end{minipage}
    \hfill
    \begin{minipage}[t]{0.31\linewidth}
        \centering
        \includegraphics[width=\linewidth]{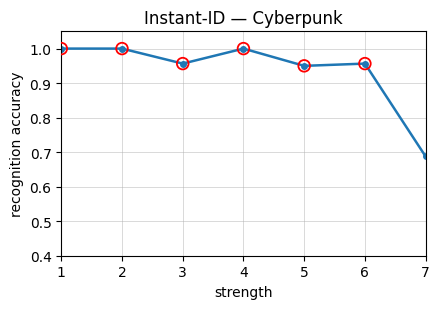}
    \end{minipage}\vspace{-2mm}
    \caption{Recognition accuracy as a function of stylization strength on StyleBench-S. The x-axis denotes stylization strength and the y-axis denotes recognition accuracy. The first row shows the Pixar style across three methods, and the second row shows three different styles generated by Instant-ID.}
    \label{fig:styleid_strength}
\end{figure*}

\subsection{StyleBench-S: Large-Scale Synthetic Supervision Set}\label{subsec:StyleBench-S}
While StyleBench-H serves as a robust evaluation benchmark, training a deep identity metric requires orders of magnitude more data. Because collecting hundreds of thousands of human annotations is prohibitively expensive, we generate \textbf{StyleBench-S}, a synthetic dataset containing 220k stylized pairs explicitly calibrated to human perceptual thresholds.

To align synthetic data generation with human perception, we first derive psychometric recognition curves for each style $t$ and method $m$. These curves map stylization strength to the probability of human recognition. We conducted a calibration user study similar to StyleBench-H, using non-overlapping FFHQ source images and generating stylized images across 10 styles and 7 strength levels. Unlike StyleBench-H, this calibration study employed a Two-Alternative Forced-Choice (2AFC) protocol to estimate a retention threshold for identity across stylization methods and styles. Participants were presented with a source image $I_{src}$ and two stylized options ($I_{style1}$, $I_{style2}$), one depicting the target identity and the other a distractor, and were asked to select the image matching the source. We applied the same quality filters as in StyleBench-H. A total of $N=76$ participants were recruited (Male: 33, Female: 37, Not disclosed: 6), with a mean age of 29.0 years. Four participants failed the consistency checks, leaving $N=72$. Each participant answered 61 queries, yielding $N_S=4315$ valid responses after latency filtering.

\begin{figure}
  \includegraphics[width=\linewidth]{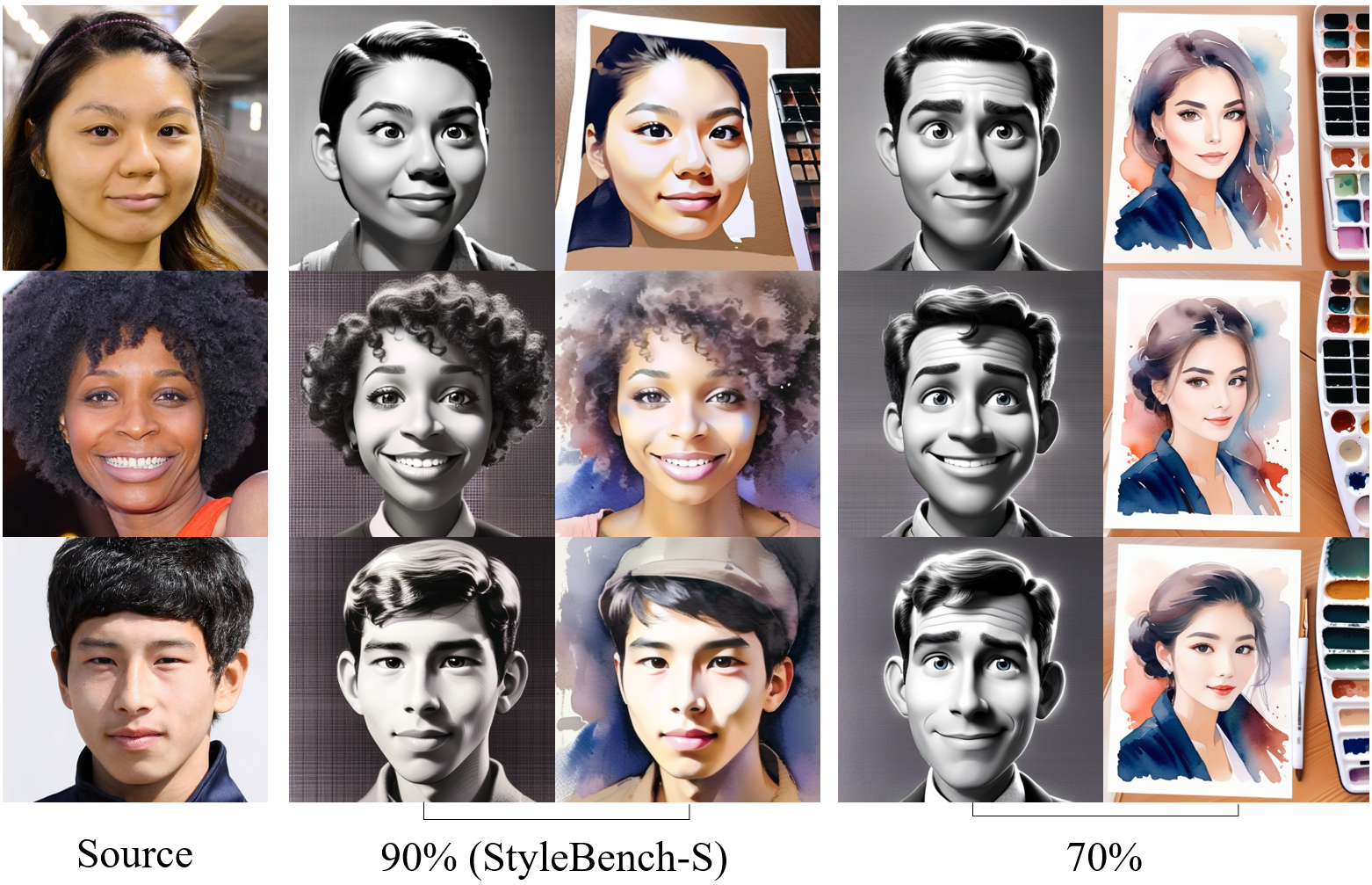}
  \vspace{-4mm}
  \caption{Comparison between StyleBench-S samples selected with a 90\% threshold and those selected with a lower threshold (70\%). The 90\% threshold preserves identity while allowing stylistic variation, whereas the lower threshold fails to maintain identity.}
  \label{fig:id-s-threshold}
\end{figure}

Using these responses, we plot psychometric functions of recognition accuracy versus stylization strength for each method (Figure~\ref{fig:styleid_strength}). The resulting curves vary substantially across stylization frameworks (top row) and across artistic styles within the same framework (bottom row). In particular, IP-Adapter shows a sharp drop in recognition accuracy as stylization strength increases, while InstantID and InfiniteYou exhibit more gradual degradation, indicating stronger inherent identity retention. Moreover, even under the same discretization, different artistic styles induce markedly different perceptual behaviors, confirming that identity preservation cannot be characterized by a single global threshold. These findings motivate method- and style-aware calibration when constructing synthetic supervision aligned with human perception.

Based on these findings, we leverage the psychometric curves to generate our large-scale synthetic dataset, StyleBench-S. Specifically, we select stylized image pairs at strength levels where the human recognition probability remains high (e.g., above 90\%) and designate them as perceptual positives according to the calibrated thresholds. As illustrated in Figure~\ref{fig:id-s-threshold}, a 90\% threshold reliably preserves identity-consistent features aligned with human judgment, whereas relaxing the threshold to 70\% results in noticeable identity degradation. This discrepancy arises because, in the 2AFC setting, observers can often rely on coarse semantic cues (e.g., gender) to make correct decisions even at lower recognition levels. While such cues are sufficient for binary discrimination, they are overly permissive for the fine-grained identity preservation we aim to model. Therefore, among perceptual positive candidates, we retain only samples corresponding to the highest and second-highest recognition levels for each method--style combination to construct StyleBench-S. The complete set of psychometric curves is presented in Figure~\ref{fig:id-s-full}.

\paragraph{Dataset configuration.}
StyleBench-S is organized by identity, where each identity corresponds to a unique source subject. The final dataset contains 4,073 identities, each associated with 55 stylized images generated from the same source identity under multiple method and style configurations that satisfy the calibrated perceptual thresholds. Specifically, we generate images across 10 artistic styles for each of three stylization frameworks, yielding 30 distinct style outputs per identity. When multiple high-confidence strength levels are available, we additionally include variants from the top two perceptually reliable levels. Together, this procedure constructs a total of 55 stylized images per identity, resulting in approximately 224k stylized samples overall.
\section{Stylization-Agnostic Identity Recognition Model}
\label{sec:metric}

Our next goal is to train an identity recognition model for portraits whose metrics remain robust across stylization frameworks and artistic styles. To this end, we introduce StyleID, a perception-calibrated identity encoder that maps an image to an identity embedding and is trained to match human judgments under controlled stylization. We build StyleID on top of CLIP~\cite{radford2021clip}, a large-scale text–image semantic encoder, because its representations are comparatively robust to appearance shifts and texture/style variations, providing a strong foundation for identity modeling under out-of-distribution stylizations.

We use StyleBench-S as training supervision to adapt CLIP to stylized identity recognition while keeping the pretrained backbone fixed. Concretely, we freeze the CLIP image encoder and inject LoRA adapters into attention and linear layers to learn a lightweight, style-robust identity representation without overfitting or catastrophic drift from CLIP’s pretrained manifold. On top of the resulting embedding, we attach an angular margin function similar to ArcFace~\cite{deng2019arcface} that enforces discriminative angular margins between identities, enabling consistent separation even under large stylistic shifts.

\begin{figure}[t]
\centering
  \includegraphics[width=\linewidth]{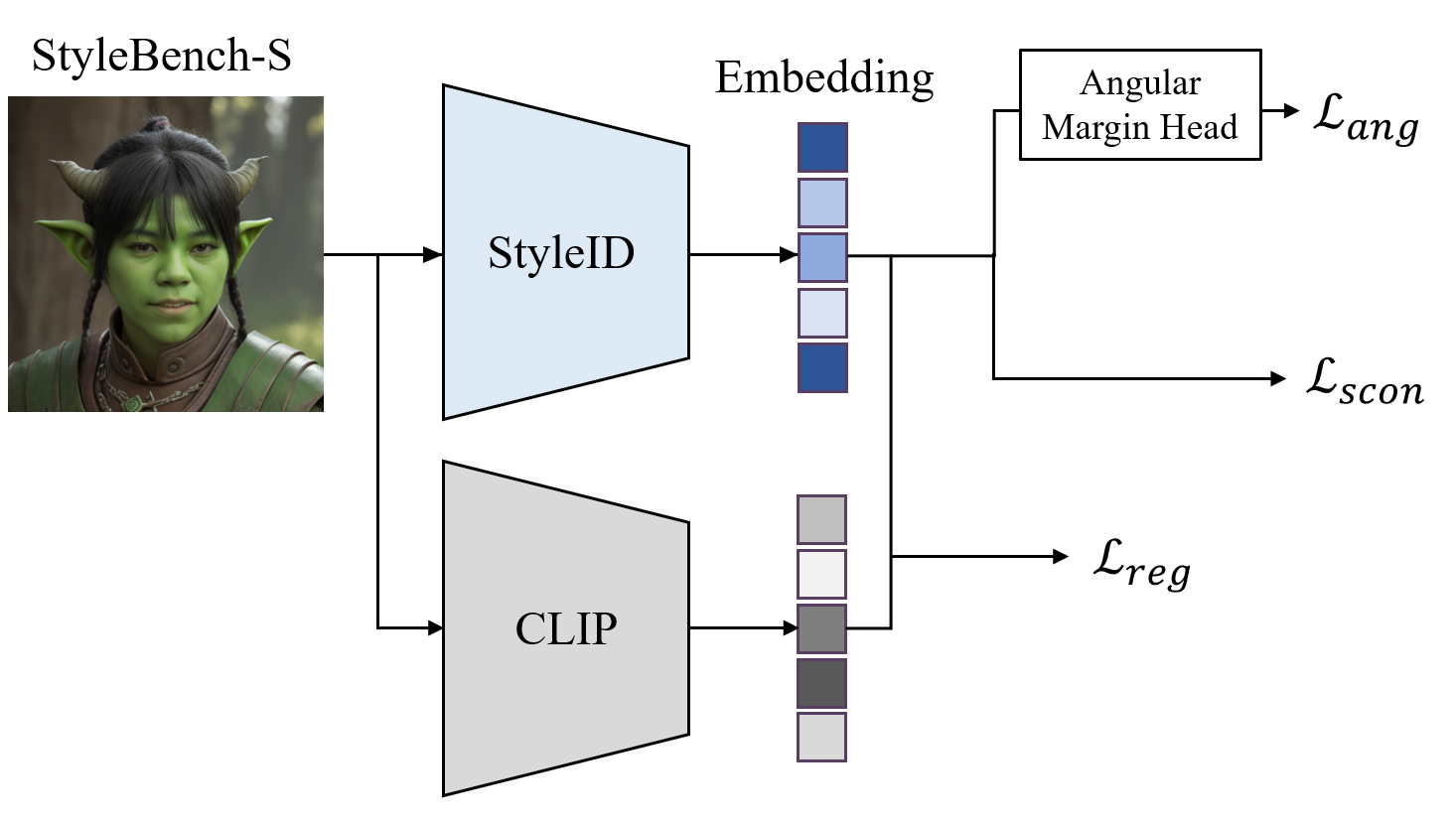}\vspace{-3mm}
  \caption{StyleID training overview. While the angular margin loss is computed with respect to class centers and the supervised contrastive loss is computed with \fixq{positive-negative} pairs rather than individual samples, we visualize it using sample-level relationships for simplicity.}
  \label{fig:train}
\end{figure}

Our training objective combines three terms: an angular identity loss, a supervised contrastive loss, and an embedding regularization loss that stabilizes adaptation as presented in Figure~\ref{fig:train}. For the angular identity loss, we adopt the standard ArcFace loss. Given an embedding $\mathbf{z}_i \in \mathbb{R}^d$ for sample $i$, we $\ell_2$-normalize it as $\hat{\mathbf{z}}_i=\mathbf{z}_i/\|\mathbf{z}_i\|_2$. The angular head maintains class weights $\mathbf{w}_c$, which are also normalized as $\hat{\mathbf{w}}_c=\mathbf{w}_c/\|\mathbf{w}_c\|_2$. The cosine logit for class $c$ is:
\begin{equation}
\cos\theta_{i,c} = \hat{\mathbf{z}}_i^\top \hat{\mathbf{w}}_c. \qquad
\theta_{i,c}=\arccos(\cos\theta_{i,c}) \in [0,\pi].
\end{equation}
With additive angular margin $m$ and scale $\alpha$, and letting $y_i$ denote the ground-truth class label of sample i, the ArcFace (cross-entropy) loss for sample $i$ is computed as follows:
\begin{equation}
\ell_i^{\text{ang}}
= -\log
\frac{
\exp\!\left(\alpha \cdot \cos(\theta_{i,y_i}+m)\right)
}{
\exp\!\left(\alpha \cdot \cos(\theta_{i,y_i}+m)\right)
+
\sum_{c\neq y_i} \exp\!\left(\alpha \cdot \cos\theta_{i,c}\right)
}, \label{eq:angle}
\end{equation}
and the batch loss is as follows where $B$ indicates the size of the minibatch:
\begin{equation}
\mathcal{L}_{\text{ang}}=\frac{1}{B}\sum_{i=1}^{B}\ell_i^{\text{ang}}.
\end{equation}

In addition to angular identity loss, for better separation between the different identities, we compute a supervised contrastive loss~\cite{khosla2020supervised} on the normalized embeddings $\{\hat{\mathbf{z}}_i\}_{i=1}^{B}$, which explicitly pulls together samples of the same identity while pushing apart different identities at the instance level, complementing the class-level margin enforced by $\mathcal{L}_{\text{ang}}$. For each anchor $i$, positives are defined as samples in the minibatch sharing the same identity:
\begin{equation}
\mathcal{P}(i)=\{\,p \in \{1,\dots,B\}\setminus\{i\}\;|\; y_p = y_i \,\}.
\end{equation}
Using temperature $\tau$, the supervised contrastive loss for anchor $i$ is
\begin{equation}
\ell_i^{\text{scon}}
=
-\frac{1}{|\mathcal{P}(i)|} \; 
\sum_{p\in \mathcal{P}(i)}
\log
\frac{
\exp\!\left(\hat{\mathbf{z}}_i^\top \hat{\mathbf{z}}_{p}/\tau\right)
}{
\sum_{a\in \{1,\dots,B\}\setminus\{i\}}
\exp\!\left(\hat{\mathbf{z}}_i^\top \hat{\mathbf{z}}_{a}/\tau\right)
}.
\end{equation}
We average across anchors:
\begin{equation}
\mathcal{L}_{\text{scon}}=\frac{1}{B}\sum_{i=1}^{B}\ell_i^{\text{scon}}.
\end{equation}

Lastly, for stable training and to prevent overfitting, we adopt an embedding regularization term that constrains the adapted representation to remain close to the frozen CLIP embedding:
\begin{equation}
\mathcal{L}_{\text{reg}} = \frac{1}{B}\sum_{i=1}^{B} \left\| \hat{\mathbf{z}}_i - \hat{\mathbf{z}}^{(0)}_i \right\|_2^2,
\end{equation}
where $\hat{\mathbf{z}}^{(0)}_i$ denotes the $\ell_2$-normalized embedding extracted from the original frozen CLIP encoder without LoRA, and $\hat{\mathbf{z}}_i$ is the corresponding embedding produced by the LoRA-adapted encoder. Therefore, the total training loss can be written as follows:
\begin{equation}
    \mathcal{L} = \mathcal{L}_{\text{ang}} +  \lambda_{\text{scon}} \mathcal{L}_{\text{scon}}  +  \lambda_{\text{reg}} \mathcal{L}_{\text{reg}},
\end{equation}~\label{eq:full_training}
where $\lambda_{\text{scon}}$ and $\lambda_{\text{reg}}$ are balancing weights.
\section{Experiments}
\label{sec:exp}
\subsection{Implementation Details}
We implemented StyleID using the CLIP-L~\cite{radford2021clip} vision encoder, whose architecture is based on ViT~\cite{dosovitskiy2020image}. We added LoRA~\cite{hu2022lora} layers to self-attention and linear layers, using a rank of 8 during training. For the angular loss, we set the margin $m$ and scaling factor $\alpha$ in Eq.~\eqref{eq:angle} to 0.5 and 32, respectively. Balancing weights $\lambda_\text{scon}$ and $\lambda_\text{reg}$ in Eq.~\eqref{eq:full_training} were set to 0.6 and 0.1, respectively. We trained StyleID on a single NVIDIA A6000 GPU
using an AdamW optimizer, with a learning rate of 2e-4, and a batch size of 112. Each minibatch is constructed by sampling 56 identities with two samples per identity for $\mathcal{L}_{\text{scon}}$, ensuring at least one positive per anchor. The total training was conducted for 30,000 iterations on the StyleBench-S dataset.

\begin{table*}[t]
\centering
\setlength{\tabcolsep}{1.5pt}
\caption{Comparison with baselines on \textbf{StyleBench-H}. Best results are denoted in \textbf{bold}.} \vspace{-2mm}
\resizebox{\textwidth}{!}{
\renewcommand{\arraystretch}{1.1}
\begin{tabular}{lccccc cc ccccc cc ccccc}
\toprule
& \multicolumn{5}{c}{\textbf{Cross-ID}} 
& & \multicolumn{5}{c}{\textbf{Cross-Style}}
& & \multicolumn{5}{c}{\textbf{Cross-Method}} \\
\cmidrule(lr){2-6} \cmidrule(lr){8-12} \cmidrule(lr){14-18}
Methods
& TPR$\uparrow$ & Acc@0.3$\uparrow$ & Acc@0.4$\uparrow$ & Acc@0.5$\uparrow$ & AUROC$\uparrow$
&& TPR$\uparrow$ & Acc@0.3$\uparrow$ & Acc@0.4$\uparrow$ & Acc@0.5$\uparrow$ & AUROC$\uparrow$
&& TPR$\uparrow$ & Acc@0.3$\uparrow$ & Acc@0.4$\uparrow$ & Acc@0.5$\uparrow$ & AUROC$\uparrow$ \\
\midrule
ArcFace
& 0.7649 & 0.7925 & 0.7266 & 0.6449 & 0.9418
&& 0.8511  & 0.8511 & 0.7739 & 0.6729 & 0.9690
&& 0.3721 & 0.5031 & 0.5000 & 0.5000 & 0.8697 \\
AdaFace
& 0.7835 & 0.8175 & 0.7719 & 0.6849 & 0.9498
&& 0.8563 & 0.8670 & 0.8189 & 0.7101 & 0.9758
&& 0.3170 & 0.5120 & 0.5000 & 0.5000 & 0.8706 \\
\midrule
CLIP
& 0.2560 & 0.5083 & 0.5652 & 0.7015 & 0.8122
&& 0.5213  & 0.5160 & 0.5957 & 0.7660 & 0.8845
&& 0.2127 & 0.5102 & 0.5736 & 0.7362 & 0.8280 \\
SigLIP2
& 0.1736 & 0.4996 & 0.5077 & 0.5703 & 0.8119
&& 0.3245 & 0.5000 & 0.5000 & 0.5399 & 0.8488
&& 0.1431 & 0.5000 & 0.5010 & 0.5072 & 0.8609 \\
\midrule
StylizedFace
& 0.8878 & 0.9189 & 0.8837 & 0.8471 & \textbf{0.9770}
&& 0.8617 & 0.9229 & 0.8936 & 0.8697 & 0.9627
&& 0.5030 & 0.6217 & 0.5266 & 0.5000 & 0.9093 \\
\midrule
StyleID
& \textbf{0.9020} & \textbf{0.9347} & \textbf{0.9054} & \textbf{0.8662} & 0.9711
&& \textbf{0.9255} & \textbf{0.9521} & \textbf{0.9362} & \textbf{0.9149} & \textbf{0.9843} 
&& \textbf{0.7444}  & \textbf{0.8272} & \textbf{0.7658} & \textbf{0.7117} & \textbf{0.9653} \\
\bottomrule
\end{tabular}
}
\label{tab:stylebenchh_splits}
\end{table*}

\begin{table}[t]
\centering
\setlength{\tabcolsep}{3.5pt}
\caption{Comparison with baselines on \textbf{SKSF-A}. Best results are denoted in \textbf{bold}.} \vspace{-2mm}
\resizebox{\linewidth}{!}{
\begin{tabular}{lccccc}
\toprule
Methods 
& TPR$\uparrow$ & Acc@0.3$\uparrow$ & Acc@0.4$\uparrow$ & Acc@0.5$\uparrow$ & AUROC$\uparrow$ \\
\midrule
ArcFace  
& 0.6189 & 0.6801 & 0.5346 & 0.5059 & 0.9360 \\
AdaFace 
& 0.6993 & 0.6178 & 0.5357 & 0.5032 & 0.9582 \\ 
\midrule
CLIP  
& 0.4968 & 0.5000 & 0.5059 & 0.5778 & 0.9420 \\
SigLIP2 
& 0.1588 & 0.5000 & 0.5000 & 0.5011 & 0.8563 \\
\midrule
StylizedFace
& 0.4840 & 0.6279 & 0.5373 & 0.5027 & 0.9163 \\
\midrule
StyleID  
& \textbf{0.8891} & \textbf{0.8731} & \textbf{0.7393} & \textbf{0.6178} & \textbf{0.9922} \\
\bottomrule
\end{tabular}
}
\label{tab:sksfa_results}
\end{table}

\subsection{Comparison with Baselines}
\label{subsec:baseline}

\subsubsection{Baseline Methods}
We compared our method against four widely used models and one specialized model for stylized face recognition. These include two identity-focused face recognition models, ArcFace~\cite{deng2019arcface} and AdaFace~\cite{kim2022adaface} pretrained on MS1MV2~\cite{guo2016ms}, as well as two semantic image–text representation models, CLIP~\cite{radford2021clip} and SigLIP2~\cite{tschannen2025siglip}. This experiment establishes a strong and practically relevant baseline, as these encoders are frequently repurposed both as identity-preservation evaluators and as training objectives within modern face stylization and avatar generation systems, despite being originally trained and calibrated on natural photographic data (ArcFace and AdaFace) or general images (CLIP and SigLIP2). In addition, we include StylizedFace~\cite{peng2025stylized}, which is most closely related to our work. Because their code and dataset are not publicly available, we reimplemented their method based on the descriptions in the paper and trained it using StyleBench-S.

ArcFace is one of the most widely adopted face recognition models and serves as a canonical angular-margin–based identity encoder. AdaFace extends ArcFace by introducing an adaptive margin conditioned on feature norm as a proxy for sample quality. In contrast, CLIP and SigLIP2 are large-scale semantic representation models trained on image–text pairs, and are increasingly used to assess perceptual similarity and semantic consistency under significant appearance changes. StylizedFace, by comparison, is specifically designed for stylized face recognition and therefore provides a task-aligned baseline beyond general-purpose identity or semantic encoders. Together, these baselines span both identity-specific and semantic similarity paradigms, reflecting the metrics most commonly used in practice~\cite{gal2022stylegan,yoon2024lego,lee2025stylemm,chong2022jojogan,wang2024instantid} for evaluating identity preservation under stylization.

\subsubsection{Evaluation Sets}
For comparison, we conduct experiments on two complementary datasets: StyleBench-H, a human-annotated benchmark designed to capture perceptual identity judgments under controlled face stylization, and SKSF-A~\cite{yun2024stylized}, an artist-drawn sketch dataset comprising high-quality portraits across seven distinct artistic styles for each identity, representing one of the most diverse artist-drawn paired datasets in terms of style variety. Note that there is no identity overlap among StyleBench-S, StyleBench-H, and SKSF-A. StyleBench-H enables evaluation against human consensus on identity consistency across varying stylization strengths, as well as generalization to unseen identities, styles, and methods, while SKSF-A provides a challenging out-of-distribution testbed with substantial geometric and textural abstraction, allowing us to assess the robustness of identity metrics under extreme artistic transformations.

\subsubsection{Evaluation Protocols}
We evaluate identity preservation using three complementary verification metrics: the true positive rate (TPR) at a fixed false positive rate (FPR), verification accuracy, and the area under the receiver operating characteristic curve (AUROC). TPR at a fixed FPR measures the proportion of correctly verified same-identity pairs under a constrained false acceptance rate, reflecting performance in security-sensitive or high-precision settings where false positives must be tightly controlled. Here, we report TPR at FPR = $10^{-2}$. Verification accuracy measures the proportion of correctly classified positive and negative pairs under a fixed operating threshold, providing an intuitive summary of overall verification performance. We use a cosine similarity threshold in the range of 0.3–0.5, which is widely adopted in identity verification settings. For baseline comparisons, we report accuracy across all thresholds in the 0.3–0.5 range, whereas for other experiments we use a single threshold of 0.4 for simplicity unless otherwise noted. AUROC summarizes the full trade-off between TPR and FPR across all possible thresholds, offering a threshold-independent measure of discriminability between same-identity and different-identity pairs. Together, these metrics capture both operating-point–specific behavior and the global separability of identity embeddings under stylization.

\subsubsection{Comparison Results}
On StyleBench-H, the semantic encoders (CLIP and SigLIP2) exhibit the weakest agreement with human identity judgments as shown in Table~\ref{tab:stylebenchh_splits}, reflecting their training objective on broad semantic alignment rather than fine-grained identity preservation. Identity-focused face recognition models (ArcFace and AdaFace) improve substantially over semantic baselines, yet perform unsatisfactorily under stylization. Specifically, CLIP and SigLIP2 achieved only $0.17$--$0.25$ TPR, while ArcFace and AdaFace reached $0.76$--$0.78$, which remained insufficient for reliable verification under appearance shifts. StylizedFace performed substantially better than the general-purpose baselines, confirming the value of training specifically for stylized face recognition, using StyleBench-S. However, its performance remained consistently below that of StyleID, particularly in the more challenging Cross-Style and Cross-Method splits. In contrast, StyleID consistently demonstrated the best performance across nearly all reported metrics, exceeding $0.9$ TPR in Cross-ID and Cross-Style and yielding higher verification accuracy across all thresholds in the 0.3--0.5 range.

StyleID also achieved superior AUROC values, indicating strong separability of same-identity and different-identity pairs despite stylistic transformations. These results highlight the need for an identity encoder that is explicitly calibrated to be style-robust and human-aligned. We observed the same trend on the artist-drawn SKSF-A dataset (Table~\ref{tab:sksfa_results}), where StyleID maintained robust verification performance under more extreme, out-of-distribution abstractions, whereas the performance of most baselines degraded. Although StylizedFace is trained for stylized face recognition, its performance on SKSF-A remained limited, suggesting weaker generalization to artist-drawn sketches outside its training setting. CLIP improved marginally, but still remained insufficient for reliable verification.

\begin{figure*}[t]
\vspace{1mm}
    \hspace*{-8mm}
    \begin{minipage}[t]{0.24\linewidth}
        \centering
        \includegraphics[width=1.07\linewidth]{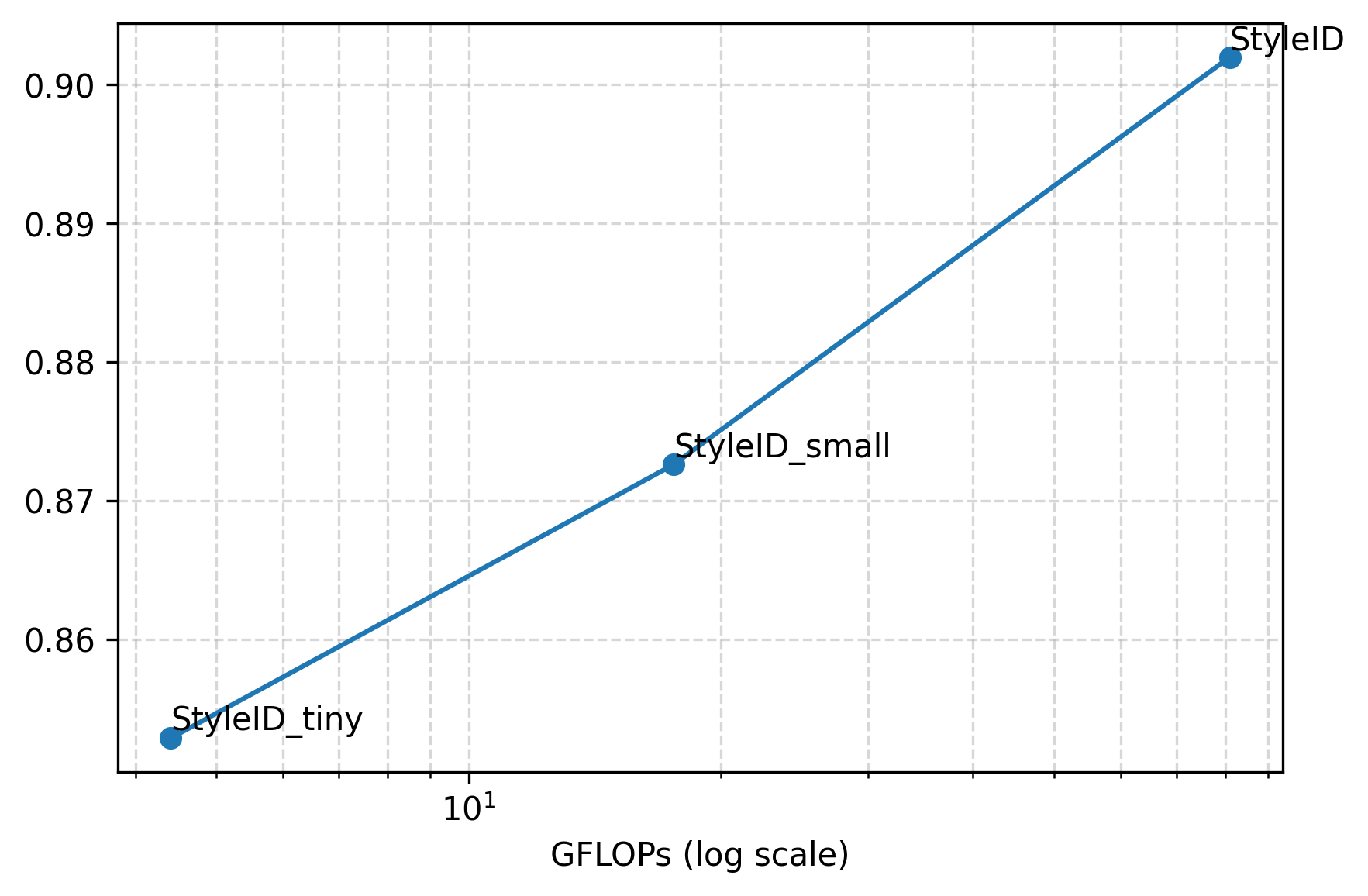}
        \vspace{-6mm}
        \caption*{(a) StyleBench-H (TPR)}
    \end{minipage}
    \begin{minipage}[t]{0.24\linewidth}
        \centering
        \includegraphics[width=1.07\linewidth]{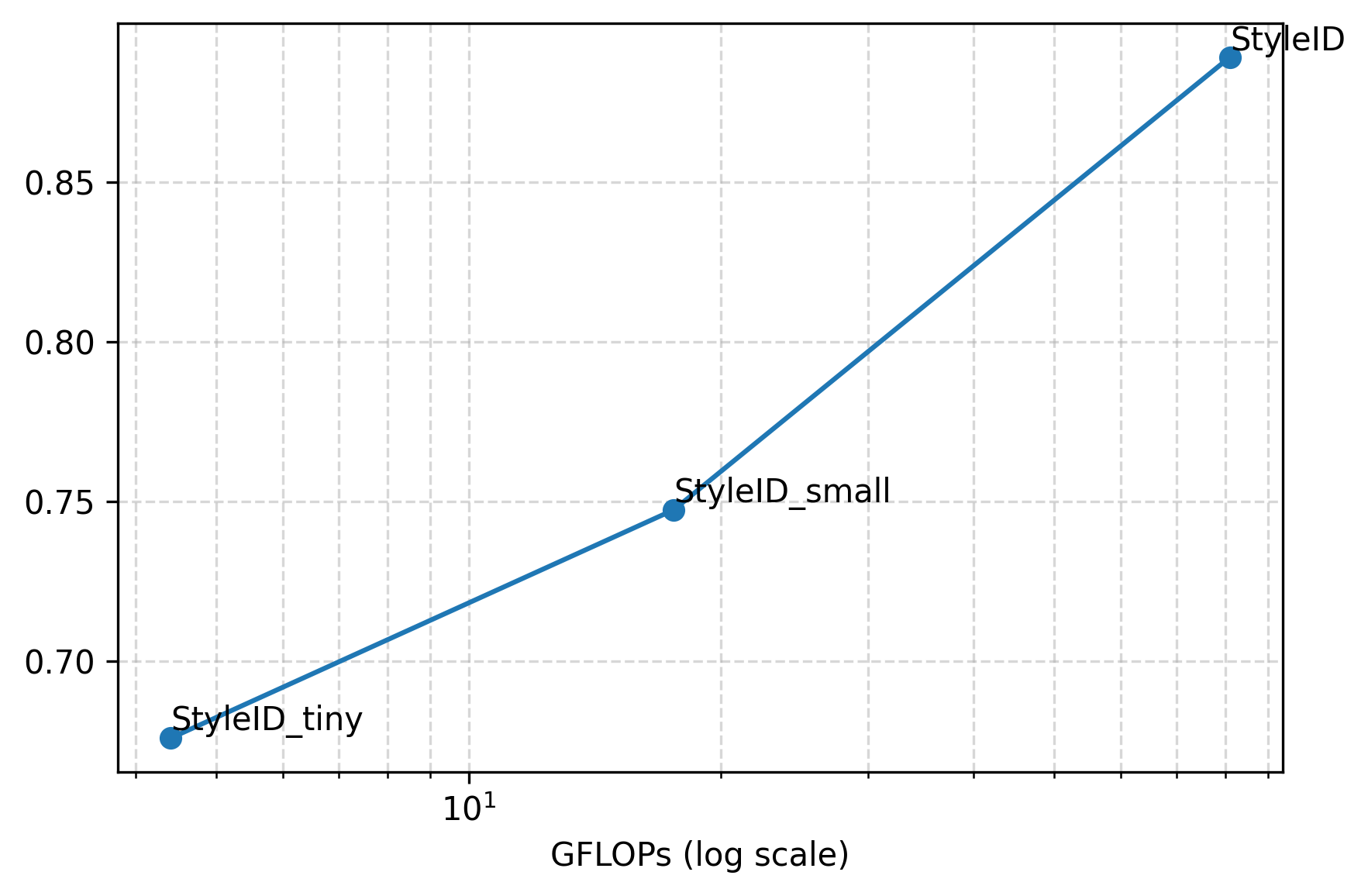}
        \vspace{-6mm}
        \caption*{(b) SKSF-A (TPR)}
    \end{minipage}
    \begin{minipage}[t]{0.24\linewidth}
        \centering
        \includegraphics[width=1.07\linewidth]{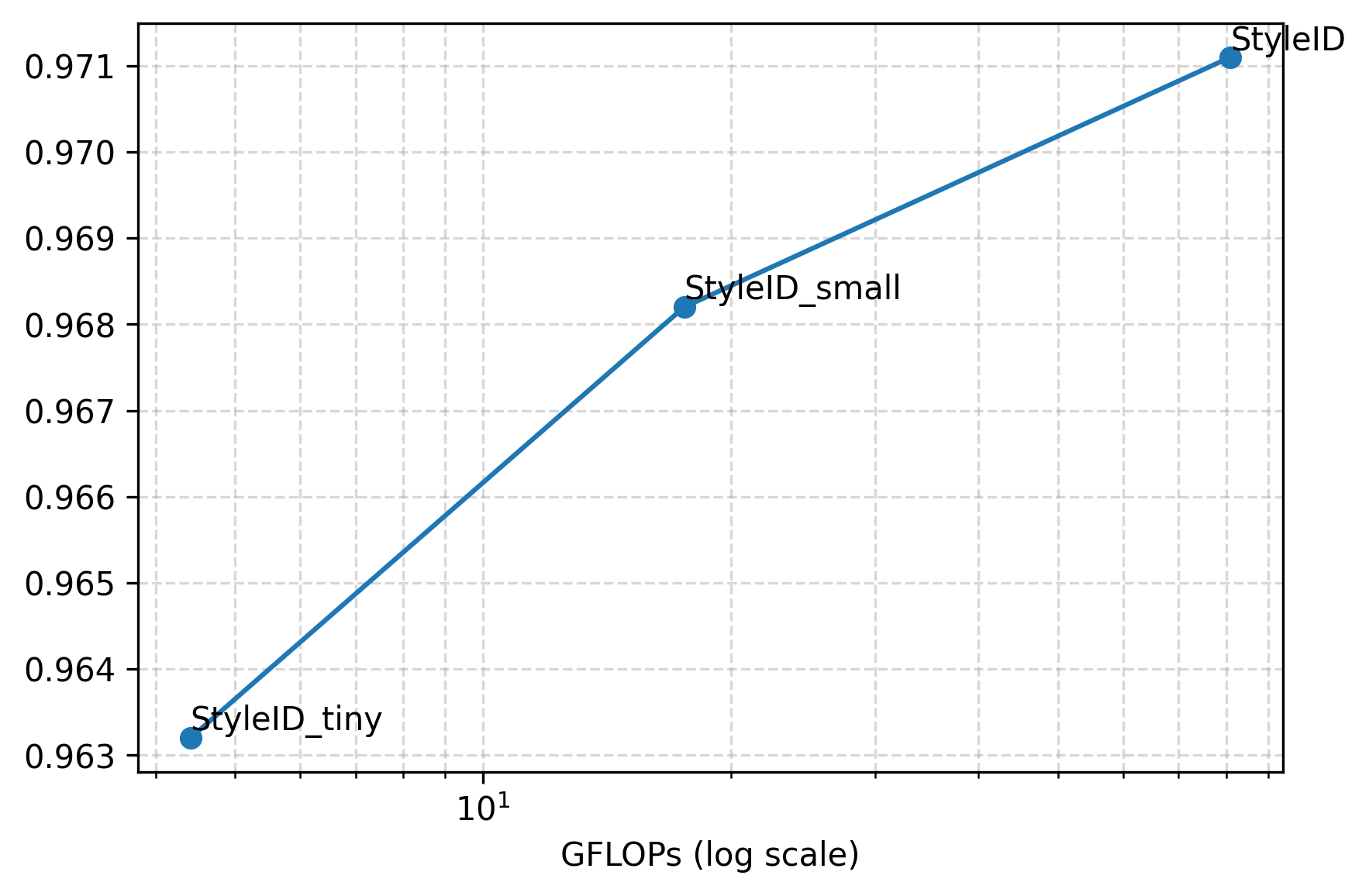}
        \vspace{-6mm}
        \caption*{(c) StyleBench-H (AUROC)}
    \end{minipage}
    \begin{minipage}[t]{0.24\linewidth}
        \centering
        \includegraphics[width=1.07\linewidth]{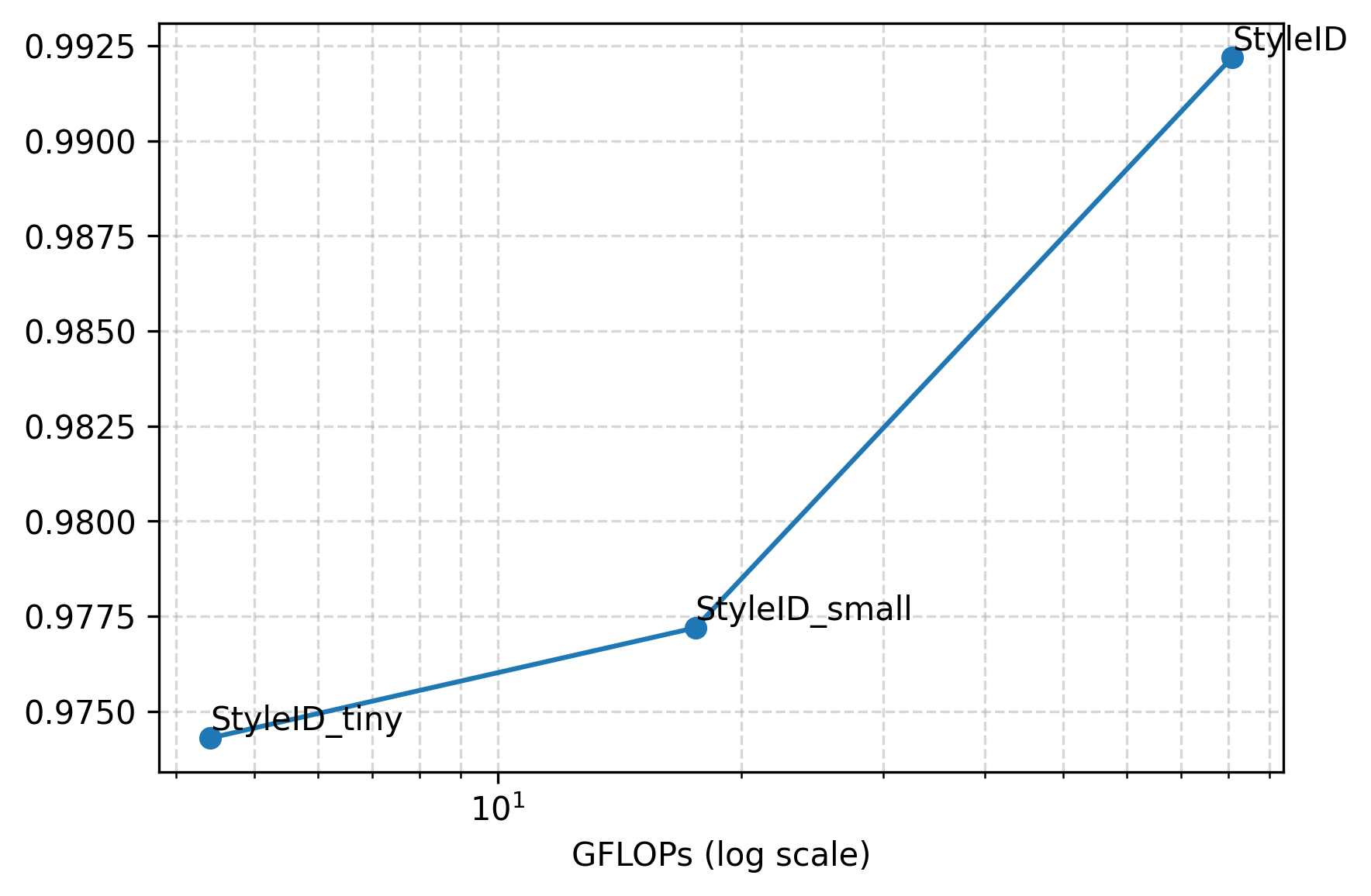}
        \vspace{-6mm}
        \caption*{(d) SKSF-A (AUROC)}
    \end{minipage}
    \caption{
    Performance versus computational cost for StyleID variants across datasets. Panels (a) and (b) show TPR versus GFLOPs, while panels (c) and (d) show AUROC versus GFLOPs. All FLOPs are plotted on a logarithmic scale. Across all datasets and metrics, performance decreases as the model size is reduced.} 
    \label{fig:scaling}
\end{figure*}

\subsection{Backbone Selection}
\label{subsec:backbone}
We adopt the CLIP-L vision encoder as the backbone of StyleID for two reasons. First, CLIP is trained on large-scale, open-vocabulary image--text data, which exposes the representation to substantial appearance diversity and makes it a strong starting point for handling the distribution shift induced by stylization. Second, CLIP provides a high-capacity ViT feature space that we can efficiently adapt with lightweight fine-tuning (LoRA) while retaining broad visual invariances. In contrast, conventional face-recognition backbones such as ArcFace and AdaFace are optimized for photographic faces and often rely on cues that are disrupted by artistic rendering, which limits their robustness under stylization.

To empirically validate this choice, we performed backbone replacement experiments. We kept the learning rate, batch size, and number of update steps identical across backbones. For ArcFace and AdaFace, which are based on an IResNet architecture~\cite{duta2021improved}, we followed a conservative adaptation protocol: we froze the network and fine-tuned only the last IResNet block and the final projection layer, which is denoted by *. This setting provides sufficient adaptation capacity while avoiding full-network retraining, which can be prone to overfitting. For comparison, $\dagger$ denotes the setting where the full parameters are trained. As reported in Table~\ref{tab:idencoder_combined}, although all methods benefited from adaptation, the gains for ArcFace* and AdaFace* were marginal, whereas SigLIP2* and our method (StyleID) improved substantially. Among these backbones, StyleID consistently achieved the best verification performance under stylization, with the highest TPR and AUROC on both datasets and the highest accuracy on StyleBench-H, while SigLIP2 attained the highest accuracy on SKSF-A. These results support our design choice of a semantically rich, high-capacity backbone as the basis for human-aligned, style-robust identity encoding.

\begin{table}[t]
\centering
\setlength{\tabcolsep}{2.5pt}
\caption{Results of backbone replacement experiments on \textbf{StyleBench-H \fixq{Cross-ID}} and \textbf{SKSF-A}. Each model is fine-tuned on StyleBench-S. Here, * denotes the conservative adaptation protocol, while $\dagger$ denotes full-parameter training. Best results are denoted in \textbf{bold}.} \vspace{-2mm}
\resizebox{\linewidth}{!}{
\begin{tabular}{lcccccc}
\toprule
& \multicolumn{3}{c}{\textbf{StyleBench-H Cross-ID}} 
& \multicolumn{3}{c}{\textbf{SKSF-A}} \\
\cmidrule(lr){2-4} \cmidrule(lr){5-7}
Methods 
& TPR$\uparrow$ & Acc$\uparrow$ & AUROC$\uparrow$
& TPR$\uparrow$ & Acc$\uparrow$ & AUROC$\uparrow$ \\
\midrule
ArcFace*  
& 0.8387 & 0.8127 & 0.9592 & 0.5597 & 0.5235 & 0.9150 \\ 
AdaFace* 
& 0.8252 & 0.8139  & 0.9535 & 0.5692 & 0.5107  & 0.9048 \\ 
ArcFace{$^\dagger$} 
& 0.8399 & 0.8161 & 0.9220 & 0.6301 & 0.5389 & 0.9284 \\ 
AdaFace{$^\dagger$}
& 0.8489 & 0.8336 & 0.9161 & 0.6013 & 0.5219 & 0.9263 \\ 
SigLIP2* 
& 0.8726 & \textbf{0.9121}  & 0.9675 & 0.8783 & \textbf{0.8443}  & 0.9862 \\ 
StyleID 
& \textbf{0.9020} & 0.9054 & \textbf{0.9711}
& \textbf{0.8891} & 0.7393  & \textbf{0.9922} \\ 
\bottomrule  \vspace{-2mm}
\end{tabular}
}
\label{tab:idencoder_combined}
\end{table}

\begin{table}[t]
\centering
\setlength{\tabcolsep}{2pt}
\caption{Ablation study results on \textbf{StyleBench-H Cross-ID} and \textbf{SKSF-A}. Best results are denoted in \textbf{bold}.} \vspace{-2mm}
\begin{tabular}{lcccccc}
\toprule
& \multicolumn{3}{c}{\textbf{StyleBench-H Cross-ID}} 
& \multicolumn{3}{c}{\textbf{SKSF-A}} \\
\cmidrule(lr){2-4} \cmidrule(lr){5-7}
Methods 
& TPR$\uparrow$ &  Acc$\uparrow$ & AUROC$\uparrow$
& TPR$\uparrow$ &  Acc$\uparrow$ & AUROC$\uparrow$ \\
\midrule
w/o $L_{ang}$  
& 0.8670 & \textbf{0.9231}  & 0.9670 & 0.8773 & \textbf{0.9302}  & 0.9921 \\
w/o $L_{scon}$
& 0.8991 & 0.9037 & 0.9704 & 0.8348 & 0.7164  & 0.9880 \\
StyleID 
& \textbf{0.9020} & 0.9054  & \textbf{0.9711}
& \textbf{0.8891} & 0.7393  & \textbf{0.9922} \\
\bottomrule
\end{tabular}
\label{tab:ablation}
\end{table}

\subsection{Ablation Study}
\label{subsec:ablation}
We conducted an ablation study on our training objectives, focusing on the angular margin loss $L_{\text{ang}}$ and the supervised contrastive loss $L_{\text{scon}}$. We evaluated each variant on StyleBench-H and SKSF-A using the same protocol as in Sec.~\ref{subsec:baseline}. The results are reported in Table ~\ref{tab:ablation}. 
Removing $L_{\text{ang}}$ led to degraded performance as shown by decreased TPR and AUROC values, because the model then relies solely on $L_{\text{scon}}$, resulting in weaker global class separation and reduced stability due to the absence of an explicit angular decision boundary. Similarly, removing $L_{\text{scon}}$ also degraded performance as shown by decreased values across all metrics, because optimizing only $L_{\text{ang}}$ is insufficient to handle the wide appearance variability introduced by stylization, which reduced robustness when style cues dominate local facial textures. Combining $L_{\text{ang}}$ and $L_{\text{scon}}$ consistently achieved the best performance in terms of TPR and AUROC, indicating that the two losses are complementary. While accuracy at a fixed threshold (0.4) is lower than w/o $L_{\text{ang}}$, this reflects threshold sensitivity rather than inferior separability, as evidenced by improved operating-point metrics (TPR) values and threshold-free metrics (AUROC) values. Overall, these results support our design choice to jointly optimize for discriminative angular margins and cross-style invariance to obtain a human-aligned, style-robust identity encoder.

\begin{figure*}[t]
\centering
  \includegraphics[width=\linewidth]{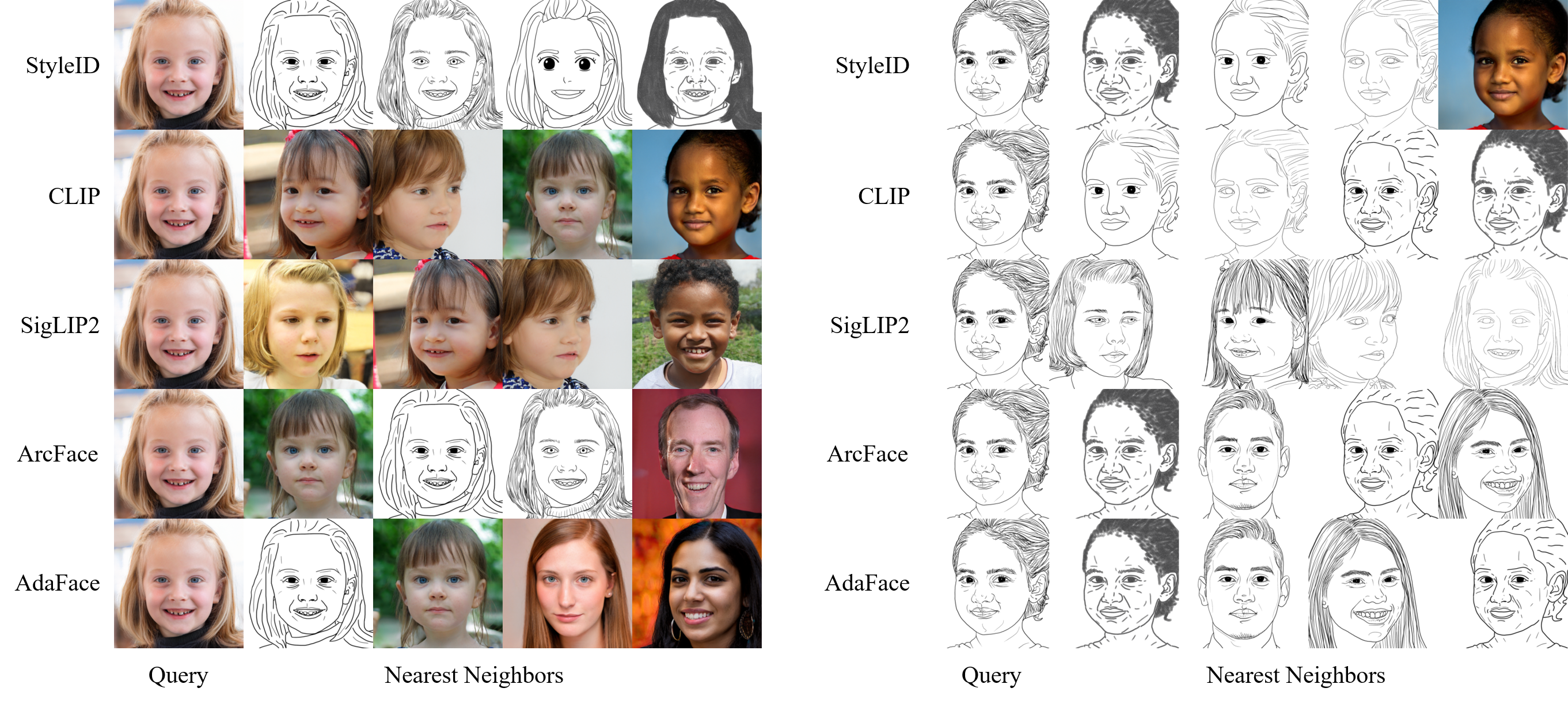}
  \caption{Retrieval results of top 4 nearest identities from \fixq{SKSF-A} dataset. Left: retrieval results from human face image. Right: retrieval results from stylized sketches.}
  \label{fig:retrieval}
\end{figure*}

\subsection{StyleID Variants}
\label{subsec:variants}
To further assess usability and computational efficiency, we evaluate lighter variants of StyleID built on smaller CLIP backbones. In addition to our default CLIP-L configuration, we instantiated StyleID with CLIP-B/16, which reduced model width and attention capacity, achieving approximately a 4$\times$ reduction in GFLOPs. This variant is denoted as StyleID\_small. We also instantiated StyleID with CLIP-B/32, which further lowered computational cost by operating on coarser $32 \times 32$ image patches, achieving approximately a 20$\times$ reduction in GFLOPs. This variant is denoted as StyleID\_tiny. As expected, reducing backbone capacity led to a gradual decrease in verification performance across datasets. Nevertheless, even the smallest variant consistently outperformed both ArcFace and AdaFace across all evaluation metrics and datasets. These lightweight variants substantially reduce memory footprint and per-image computation, making them more suitable for deployment in resource-constrained settings. This analysis examines the trade-off between backbone capacity and style-robust identity verification, and evaluates how well StyleID preserves identity under stylization as the encoder is scaled down. We report both TPR and AUROC as functions of computational cost (GFLOPs), highlighting the efficiency–accuracy trade-off of StyleID across backbone choices in Figure~\ref{fig:scaling}.

\subsection{Retrieval Test}
\label{subsec:retrieval}
We additionally evaluate each encoder as a feature extractor for identity-based retrieval. Using SKSF-A as the gallery, we computed $\ell_2$-normalized embeddings for all images and performed nearest-neighbor search based on cosine similarity. For each query, we retrieved the top-4 most similar gallery identities and assessed whether the returned samples corresponded to the same person.

As shown in Table~\ref{tab:retrieval}, StyleID achieved 84.7\% matching accuracy when queried with human face images and 70.44\% accuracy when queried with stylized sketches, outperforming the best baselines by 10–20\% in both settings. Figure~\ref{fig:retrieval} illustrates representative retrieval results for two query types: a real human face image (left) and a stylized sketch query (right). While baseline encoders often retrieved visually similar but incorrect identities under strong style variations, StyleID produced more identity-consistent neighbors across both photographic and artist-drawn queries, indicating that its embedding space better preserves identity for cross-domain retrieval.

\begin{table}[t]
\centering
\caption{Average accuracy in the retrieval test. Best results are indicated in \textbf{bold}.}
\begin{tabular}{lcc}
\toprule
Methods & \textbf{Human Query} & \textbf{Sketch Query}  \\
\midrule
ArcFace  
& 0.7313 & 0.5051 \\ 
AdaFace 
& 0.7332 & 0.5096 \\ 
CLIP
& 0.1381 & 0.6519 \\ 
SigLIP2 
& 0.0243 & 0.3782 \\ 
StyleID 
& \textbf{0.8470} & \textbf{0.7044} \\ 
\bottomrule
\end{tabular}
\label{tab:retrieval}
\end{table}

\subsection{Pose Robustness Test}
\label{subsec:pose}

To evaluate robustness under viewpoint variation, we conducted a pose consistency analysis using multi-view renderings generated with GaussianAvatar-Editor. For each identity, we rendered images across 14 viewpoints and measured cosine similarity between embeddings of different views. We report results for both realistic human images (source domain) and their stylized counterparts (target domain). This enables us to examine how identity representations behave under combined pose and style changes.

As shown in Table~\ref{tab:pose}, ArcFace exhibited strong consistency on human images, reflecting its effectiveness within the photographic domain. However, its similarity decreased under stylization, suggesting reduced stability when appearance deviates from its training distribution. In contrast, StyleID maintained more consistent similarity across both domains, leading to a higher overall average. These results indicate that while ArcFace remains a strong baseline for natural images, StyleID provides improved robustness when both pose and style variations are present.
\begin{table}[t]
\centering
\caption{Pose robustness evaluation using cosine similarity across 14 viewpoints.}
\setlength{\tabcolsep}{3pt}
\begin{tabular}{lccc}
\toprule
Method & Human (Source) & Stylized (Target) & Average \\
\midrule
ArcFace & \textbf{0.8309} & 0.4389 & 0.6349 \\
StyleID (Ours) & 0.8039 & \textbf{0.8347} & \textbf{0.8193} \\
\bottomrule
\end{tabular}
\label{tab:pose}
\end{table}
\section{Applications}
\label{sec:app}

\subsection{Enhancing Stylization}
\begin{figure}[t]
\centering
  \includegraphics[width=\linewidth]{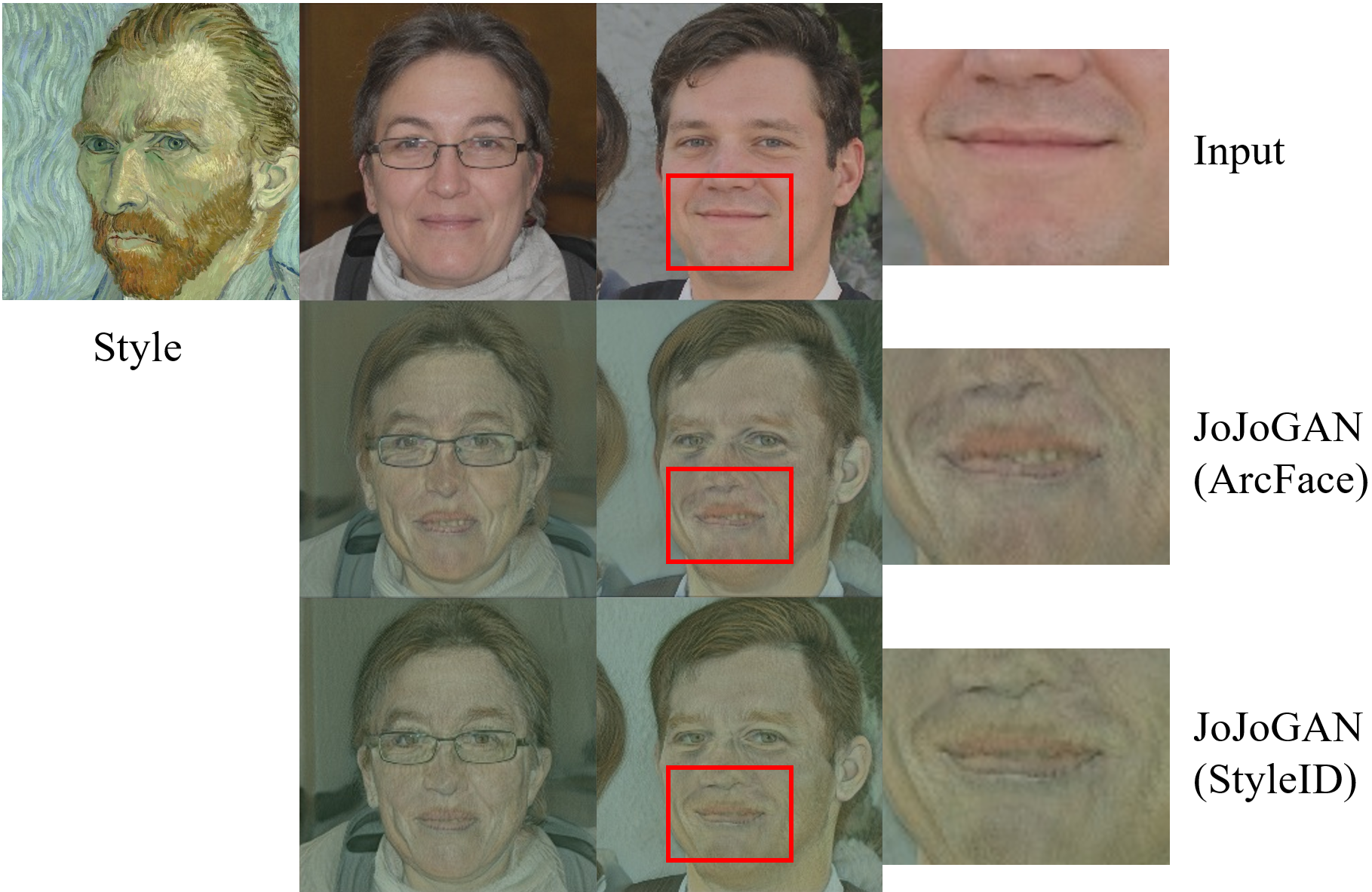}
  \caption{Application of StyleID to JoJoGAN. The original JoJoGAN using ArcFace fails to isolate style from color and introduces artifacts such as showing teeth. By adopting StyleID, the model successfully transfers the target style while maintaining the original color distribution and producing artifact-free results. \fixq{Close-up} views of the red boxes are presented on the right side of the figure.}
  \label{fig:app}
\end{figure}

We demonstrate a practical application of StyleID by replacing the identity loss in a learning-based stylization framework. Specifically, we adopt StyleID within JoJoGAN~\cite{chong2022jojogan}, which originally applies ArcFace to preserve identity during style transfer. However, because ArcFace is not designed to disentangle identity from appearance attributes such as texture and color, it often constrains stylization in undesirable ways, leading to artifacts such as residual source textures or unnatural features (e.g., incorrectly emphasized teeth) as shown in Figure~\ref{fig:app}. By substituting ArcFace with StyleID, the stylization model is guided by an identity representation that is more robust to stylistic variation. As a result, the model achieves more faithful style transfer while better preserving the original color distribution and facial structure, producing visually cleaner and artifact-free outputs. These results highlight the potential of StyleID as a drop-in replacement for conventional face recognition encoders in downstream generative applications that require identity preservation under strong appearance changes.

Following this setup, we conducted both quantitative and perceptual evaluations to compare JoJoGAN with ArcFace and StyleID as the identity constraint. For the GPT-based preference evaluation and human perceptual study, we randomly sampled 20 image pairs generated by JoJoGAN+ArcFace and JoJoGAN+StyleID. We then conducted A/B testing, asking which result was preferred in terms of identity preservation, expression preservation, and overall quality for GPT-based evaluation and additionally style preservation in the human user study. As shown in the second column of Table~\ref{tab:jojogan_style}, JoJoGAN+StyleID achieved a lower style distance (measured by Gram distance using VGG features), indicating improved style fidelity. Furthermore, automated evaluation using GPT-5.4 consistently favored StyleID in terms of identity, expression and overall quality. In addition, a user study with 15 participants (Male: 7, Female: 8) with a mean age of 27.2 years also consistently favored StyleID across style, identity, expression, and overall quality as shown in Table~\ref{tab:jojogan_pref}. These results demonstrate that StyleID provides a more suitable identity representation for stylization tasks, enabling stronger disentanglement between identity and appearance and producing outputs that are more perceptually aligned and visually coherent compared to ArcFace.

\begin{table}[t]
\centering
\caption{Quantitative evaluation of stylization fidelity and GPT-5.4 preference. Lower style distance is better; GPT scores are reported as preference rates.}
\label{tab:jojogan_style}
\resizebox{\linewidth}{!}{
\begin{tabular}{lcccc}
\toprule
Method & Style Dist $\downarrow$ & GPT-Id $\uparrow$ & GPT-Exp $\uparrow$ & GPT-Qual $\uparrow$ \\
\midrule
JoJoGAN+ArcFace & 1.398e-3 & 0.30 & 0.30 & 0.45 \\
JoJoGAN+StyleID & \textbf{1.363e-3} & \textbf{0.70} & \textbf{0.70} & \textbf{0.55} \\
\bottomrule
\end{tabular}
}
\end{table}

\begin{table}[t]
\centering
\caption{User study results. Values indicate preference rates.}
\label{tab:jojogan_pref}
\begin{tabular}{lcccc}
\toprule
Method & Style $\uparrow$ & Id $\uparrow$ & Exp $\uparrow$ & Qual $\uparrow$ \\
\midrule
JoJoGAN+ArcFace & 0.44 & 0.097 & 0.08 & 0.143 \\
JoJoGAN+StyleID & \textbf{0.56} & \textbf{0.903} & \textbf{0.92} & \textbf{0.857} \\
\bottomrule
\end{tabular}
\end{table}

\subsection{Natural-face Validation}
StyleID is primarily optimized for stylized identity evaluation rather than to replace face-recognition models on photorealistic benchmarks. To assess its transferability to natural faces, we additionally evaluated StyleID on Labeled Faces in the Wild (LFW)~\cite{huang2008labeled}. As shown in Table~\ref{tab:lfw}, StyleID remained competitive across all metrics in real-face verification, although it did not outperform ArcFace, which is specifically designed and trained for photorealistic face recognition. These results highlight that while StyleID generalizes reasonably well to natural images, its primary strength lies in modeling identity consistency under stylization rather than maximizing performance on standard face-recognition benchmarks.

\begin{table}[t]
\centering
\caption{Evaluation on the LFW benchmark for natural-face verification.}
\label{tab:lfw}
\begin{tabular}{lccc}
\toprule
Method & TPR$\uparrow$ & Acc@0.3 $\uparrow$ & AUC $\uparrow$ \\
\midrule
ArcFace & 0.9970 & 0.9975 & 0.9989 \\
StyleID & 0.9526 & 0.9750 & 0.9967 \\
\bottomrule
\end{tabular}
\end{table}
\section{Conclusion}
\label{sec:conclusion}
We presented a unified framework for perception-aligned identity evaluation and modeling under creative face stylization, consisting of two datasets and a new recognition model: StyleBench-H, StyleBench-S, and StyleID. We first introduced StyleBench-H, a human-annotated benchmark designed to evaluate identity preservation under diverse stylization methods and strengths. By explicitly controlling for stylization factors, StyleBench-H enables the collection of perception-aligned same/different identity judgments. Extensive experiments with widely used face recognition models (ArcFace, AdaFace) and semantic encoders (CLIP, SigLIP2) reveal substantial disagreement between model predictions and human judgments on stylized portraits. Furthermore, we showed that global similarity thresholds calibrated on photo-domain data fail to generalize across stylization families and intensities. These findings indicate that identity preservation under stylization remains an unsolved evaluation problem and that reliable assessment requires perception-grounded protocols.

To support scalable supervision aligned with human perception, we further propose StyleBench-S, a large-scale synthetic dataset derived from recognition-strength trends. By anchoring positive pairs with perceptual thresholds, StyleBench-S provides a structured training signal for learning identity representations that remain stable under stylization. Building on these datasets, we introduced StyleID, a new stylization-agnostic facial identity recognition model trained using StyleBench-S. Across both human-annotated and real-world stylized benchmarks, StyleID consistently achieves stronger agreement with human judgments than existing identity encoders. To improve practical usability and efficiency, we also developed lightweight variants, StyleID\_small and StyleID\_tiny, which reduce computational cost (GFLOPs) by approximately 4$\times$ to 20$\times$ compared to the full model. Together, these contributions provide both the evaluation foundation and a modeling direction for more reliable, human-aligned identity recognition in creative and generative settings.

\paragraph{Limitation and future work.}
Despite the strengths of our framework, several limitations remain.  
First, while StyleBench-H provides perception-aligned annotations across multiple stylization methods and strengths, its scale is inherently constrained by the cost of human annotation. Expanding coverage to a broader range of artistic styles, cultural aesthetics, and demographic attributes would further improve the generality of the benchmark. We also note that StyleBench-H is demographically skewed toward young, white subjects, which may bias both the benchmark and the learned metric toward identities that are overrepresented in the data. As a result, optimizing on this distribution may lead to uneven performance for underrepresented and intersectional groups, highlighting the need for broader demographic coverage and subgroup-aware evaluation in future work.

Second, StyleBench-S relies on synthetic supervision derived from recognition-strength trends from human annotation. Although this enables scalable training and leads to strong empirical performance, synthetic generation may not fully capture the complexity of real-world artistic variation. Incorporating more diverse real stylized data and exploring hybrid human–synthetic supervision are promising future directions.  

Finally, StyleID focuses on identity robustness under appearance stylization and does not explicitly model other challenging factors such as extreme pose and occlusion. This design choice stems from our data filtering strategy, which was intended to improve annotation reliability and the effectiveness of training the StyleID model. However, it may reduce robustness under extreme conditions. Extending the framework to cover broader robustness dimensions and integrating multimodal supervision remain important directions for future work.

\bibliographystyle{ACM-Reference-Format}
\bibliography{main}

\appendix

\appendix  
\section{Appendix}

In this appendix, we provide additional details and results. We organize the contents as follows:
\begin{itemize}
    \item \textbf{Section \ref{sec:appendix_dataset}}: Details on dataset construction. 
    \item \textbf{Section \ref{sec:appendix_exp}}: Additional experiment results.
\end{itemize}

\section{Dataset Construction Details}
\label{sec:appendix_dataset}

\begin{figure*}
\hspace*{-5mm}
  \includegraphics[width=1.06\linewidth]{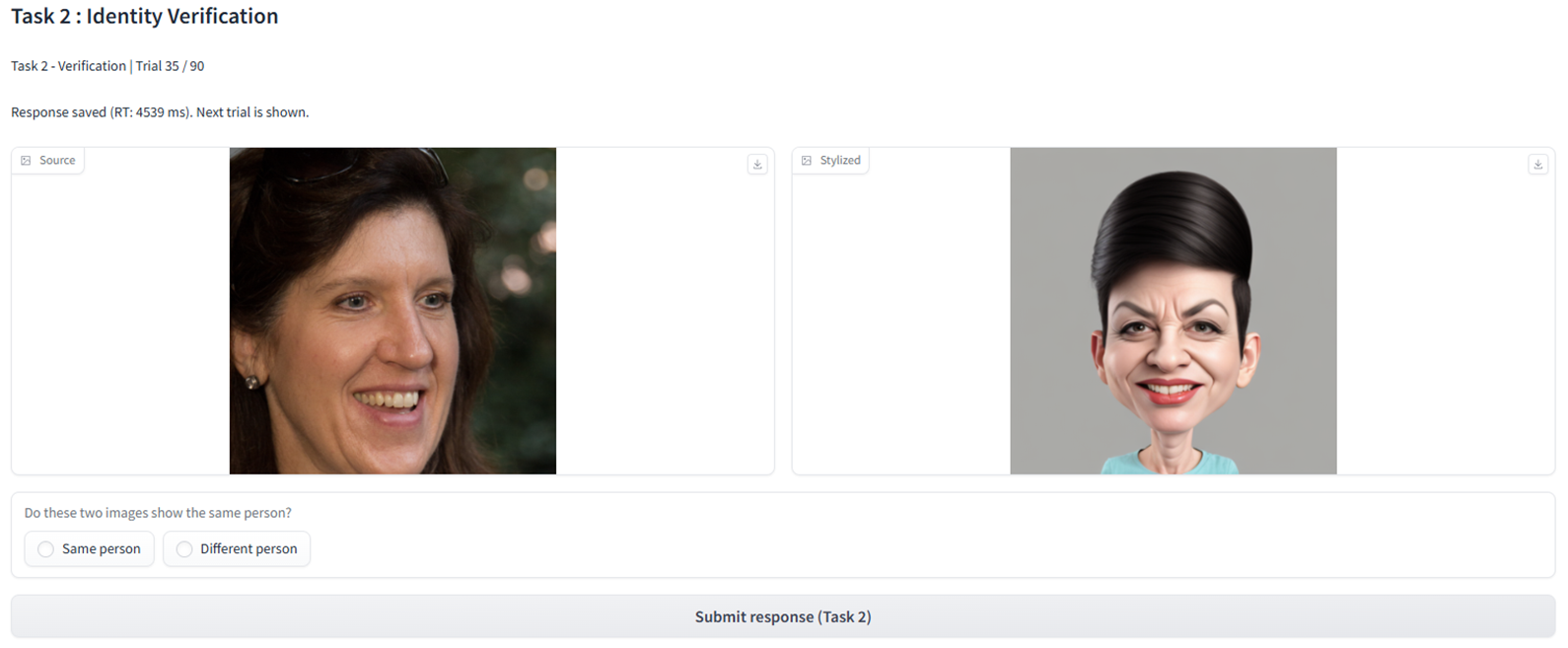} \vspace{-3mm}
  \caption{Questionnaire sample for data collection for StyleBench-H.}
  \label{fig:us-1}
\end{figure*}

\begin{figure*}
\hspace*{-5mm}
  \includegraphics[width=1.06\linewidth]{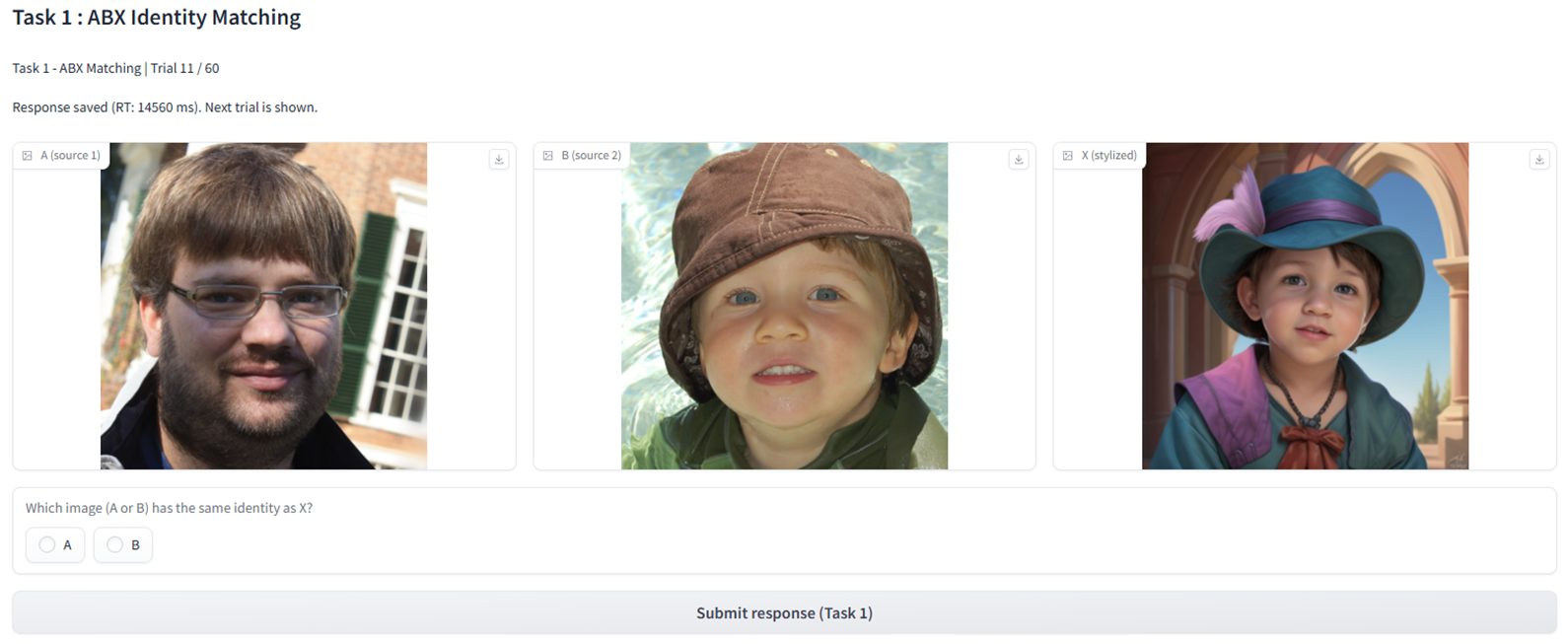} \vspace{-3mm}
  \caption{ Questionnaire samples used for data collection across different stylization methods and specific styles for StyleBench-S.}
  \label{fig:us-2}
\end{figure*}

\subsection{StyleBench-H: Human-Judged Identity Verification Benchmark}
\label{sec:appendix_styleid_h}

\textbf{Goal.} StyleBench-H is designed to measure identity preservation under controllable face stylization using human perceptual judgments. It consists of image pairs $(I_{\mathrm{src}}, I_{\mathrm{style}})$ where $I_{\mathrm{style}}$ is a stylized rendering of the source identity $I_{\mathrm{src}}$ under a chosen style and stylization strength.

\textbf{Pair generation.} For each source identity, we generate stylized counterparts across a set of artistic styles and multiple stylization methods. Each stylization is produced at discrete strength levels $s \in \{1/7, 2/7, \ldots, 7/7\}$, covering mild to extreme appearance changes. This controlled design enables analysis of recognition behavior as a function of stylization strength.

\textbf{Annotation task.} We adopt a pairwise verification protocol: annotators are shown a source photo and its stylized counterpart, and asked whether the two images depict the same person. The exact questionnaire interface is provided in Fig.~\ref{fig:us-1}. We explicitly inform annotators that the second image is a stylized rendering to discourage reliance on superficial style mismatch and to focus attention on identity-related cues such as facial structure and characteristic features.

\begin{figure*}[t]
    \centering
    \vspace{2mm}

    \begin{minipage}[t]{0.32\linewidth}
        \centering
        \includegraphics[width=\linewidth]{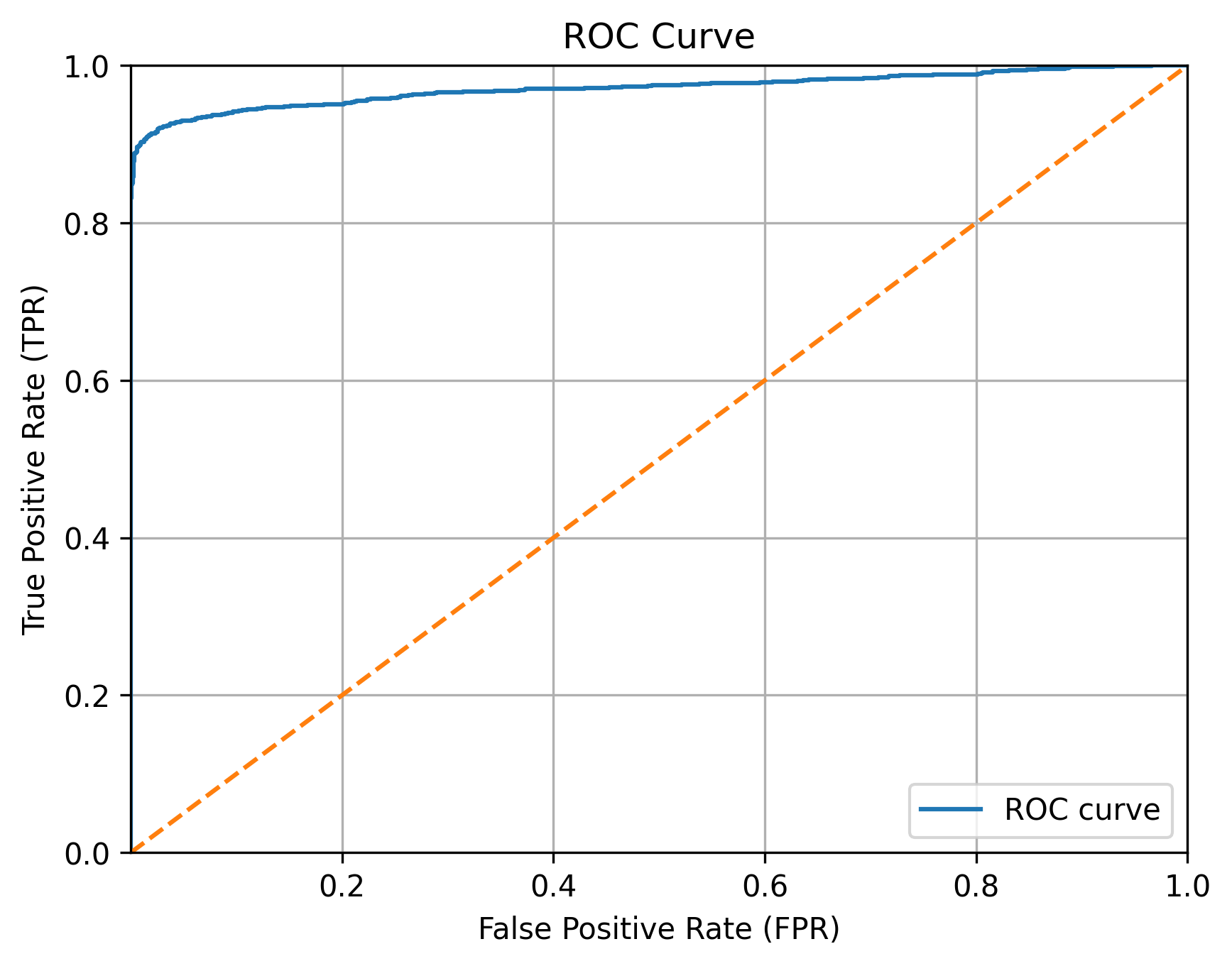}
        \vspace{-6mm}
        \caption*{(a) StyleBench-H (linear scale)}
    \end{minipage}\hspace{0.1\linewidth}
    \begin{minipage}[t]{0.32\linewidth}
        \centering
        \includegraphics[width=\linewidth]{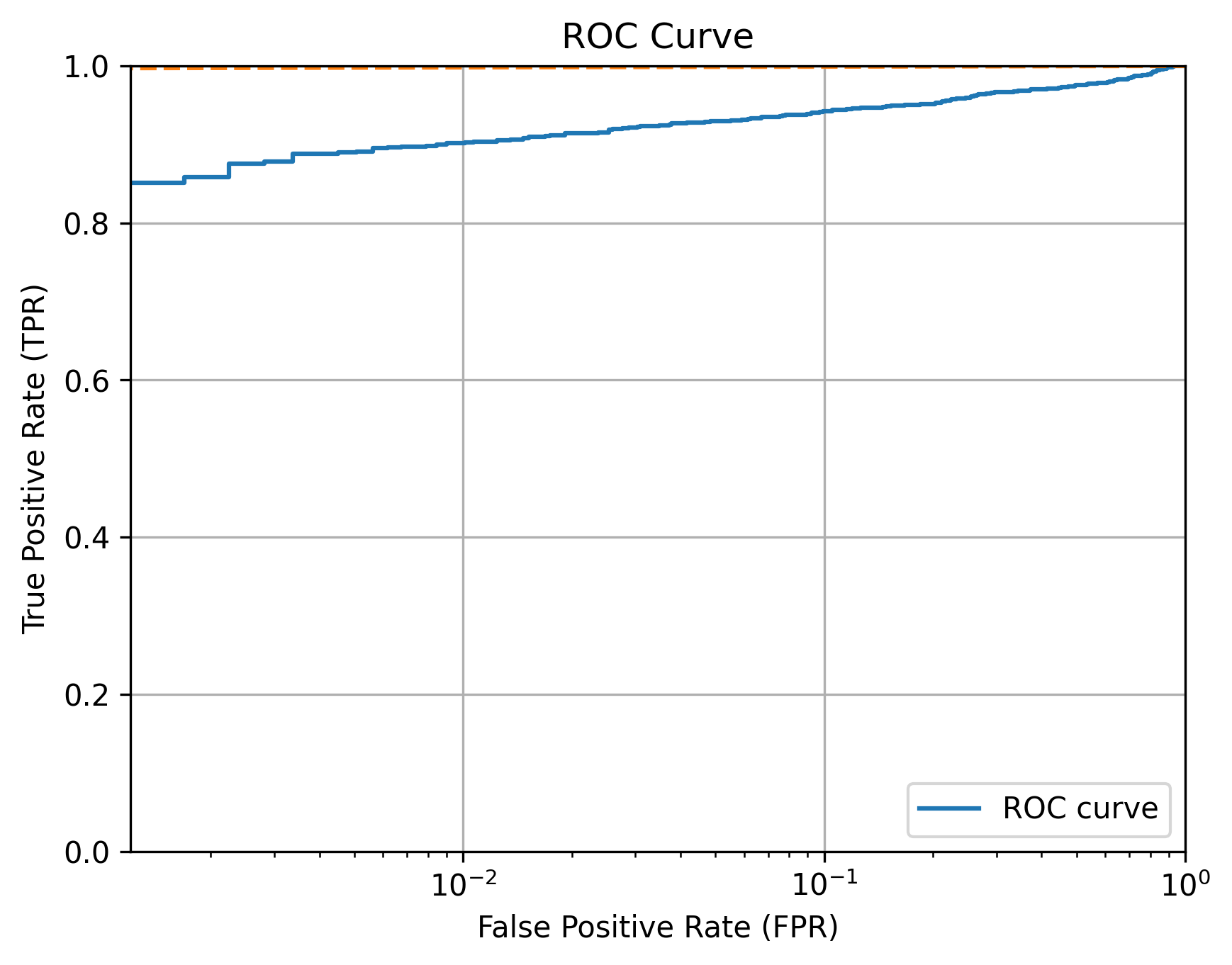}
        \vspace{-6mm}
        \caption*{(b) StyleBench-H (log scale)}
    \end{minipage}

    \vspace{3mm}

    \begin{minipage}[t]{0.32\linewidth}
        \centering
        \includegraphics[width=\linewidth]{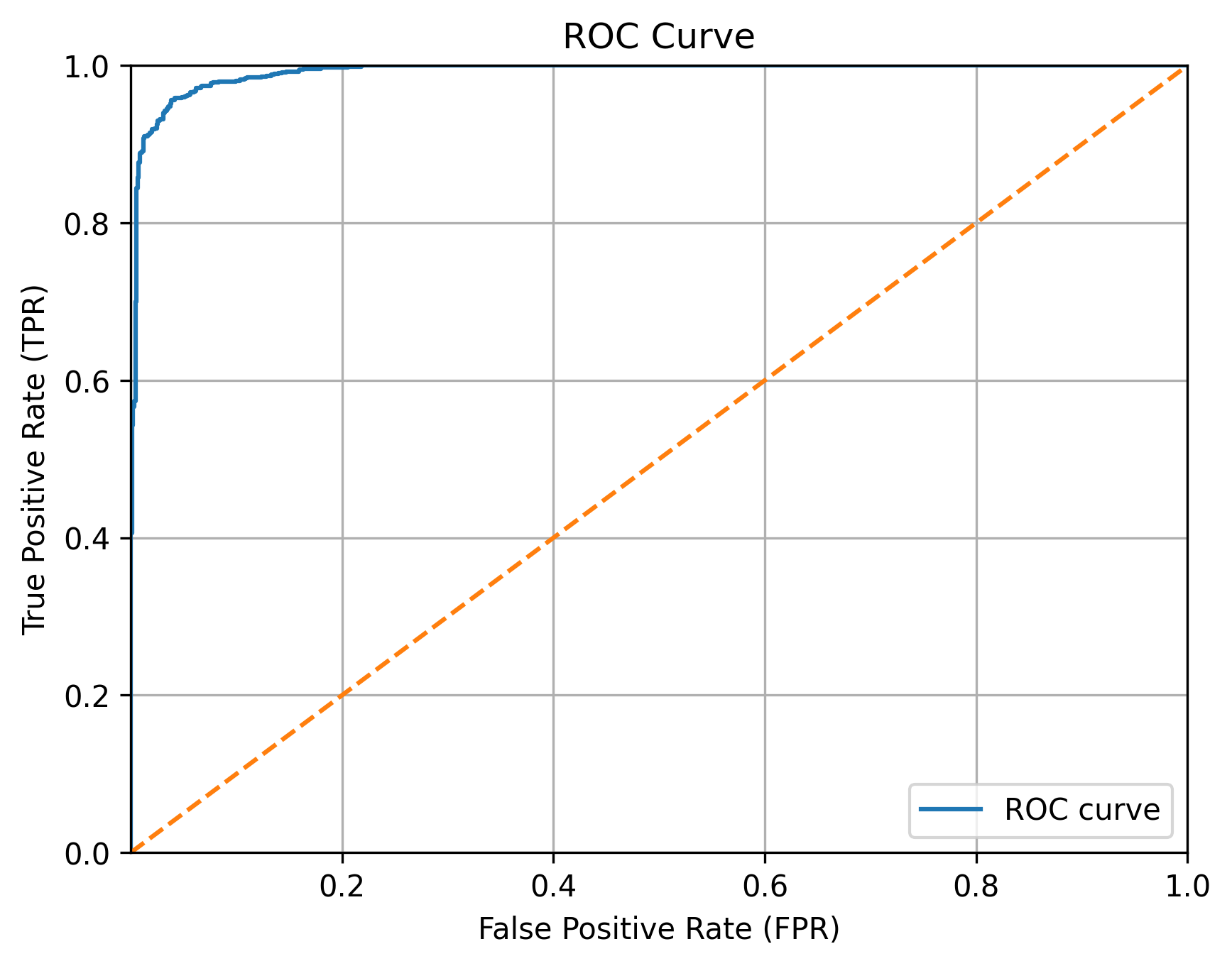}
        \vspace{-6mm}
        \caption*{(c) SKSF-A (linear scale)}
    \end{minipage}\hspace{0.1\linewidth}
    \begin{minipage}[t]{0.32\linewidth}
        \centering
        \includegraphics[width=\linewidth]{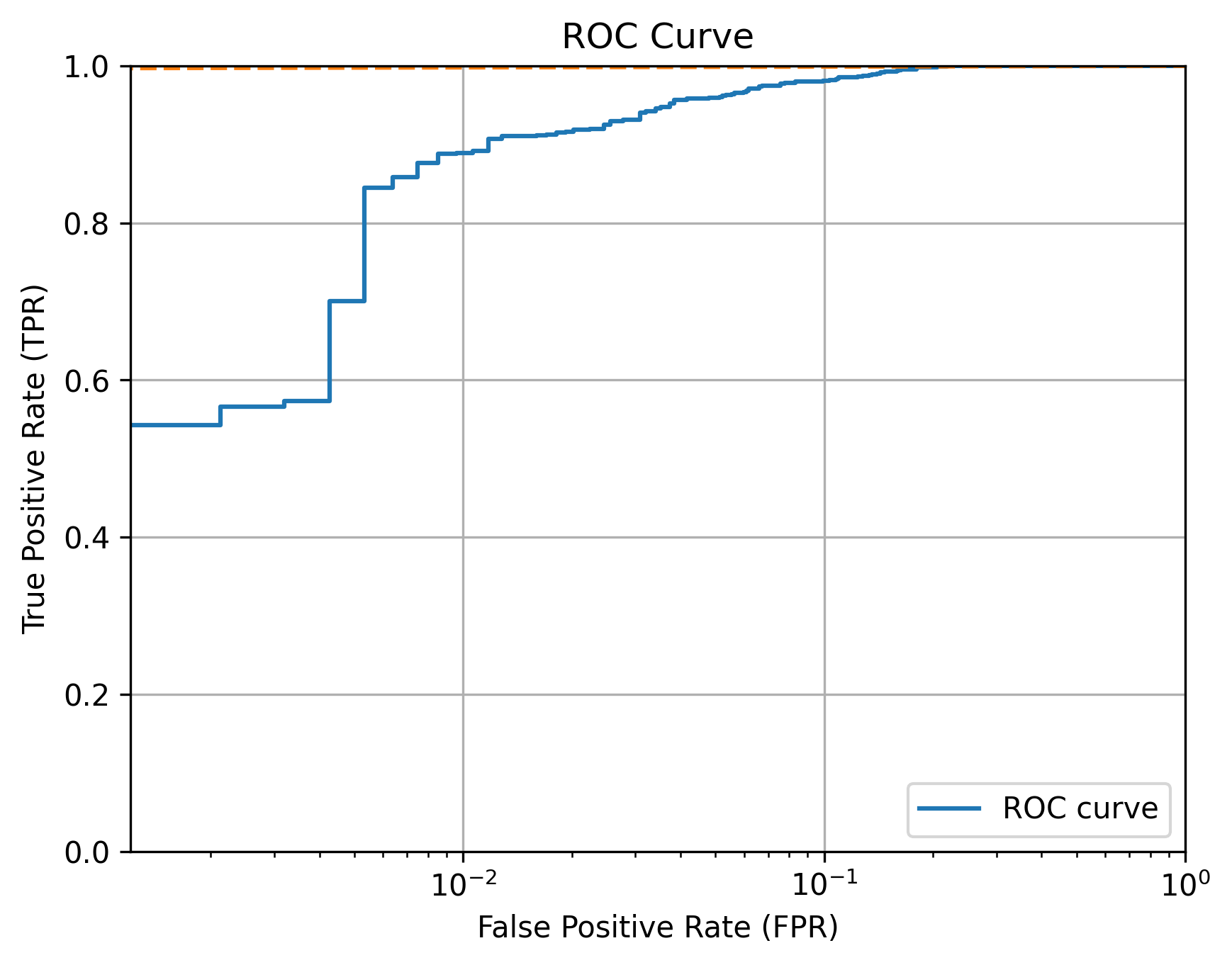}
        \vspace{-6mm}
        \caption*{(d) SKSF-A (log scale)}
    \end{minipage}

    \vspace{1mm}
    \caption{Full Receiver Operating Characteristic (ROC) curves on both linear and log scales.}
    \label{fig:roc_curve}
\end{figure*}

\subsection{StyleBench-S: Synthetic Supervision via Human-Calibrated Recognition Statistics}
\label{sec:appendix_styleid_s}
\textbf{Goal.} StyleBench-S is a large-scale synthetic training set derived from human recognition statistics. It provides scalable supervision that preserves the trends observed in StyleBench-H while enabling efficient training of a style-robust identity encoder.

\textbf{Psychometric curve construction.} For each method--style combination, we fit psychometric curves that relate stylization strength to human recognition probability. To obtain these curves, we conduct an additional human study using a structured questionnaire (Fig.~\ref{fig:us-2}) and compute recognition accuracy at each strength level. We then use this recognition probability across strengths, methods, and styles.

\textbf{Selecting perceptual positives.} Using the recognition curves, we select the combination of style, method, and strength, whose estimated human recognition probability exceeds a high threshold (e.g., 90\%) and treat them as perceptual positives. This procedure filters out stylizations that are likely to degrade identity according to human judgment, producing a training signal that is both scalable and perception-aware. In practice, we further prioritize samples from the highest recognition regimes within each method--style pair to maintain strict identity consistency under stylization.

\begin{figure*}
  \vspace{2mm}\hspace{-8mm}
  \includegraphics[width=0.9\linewidth]{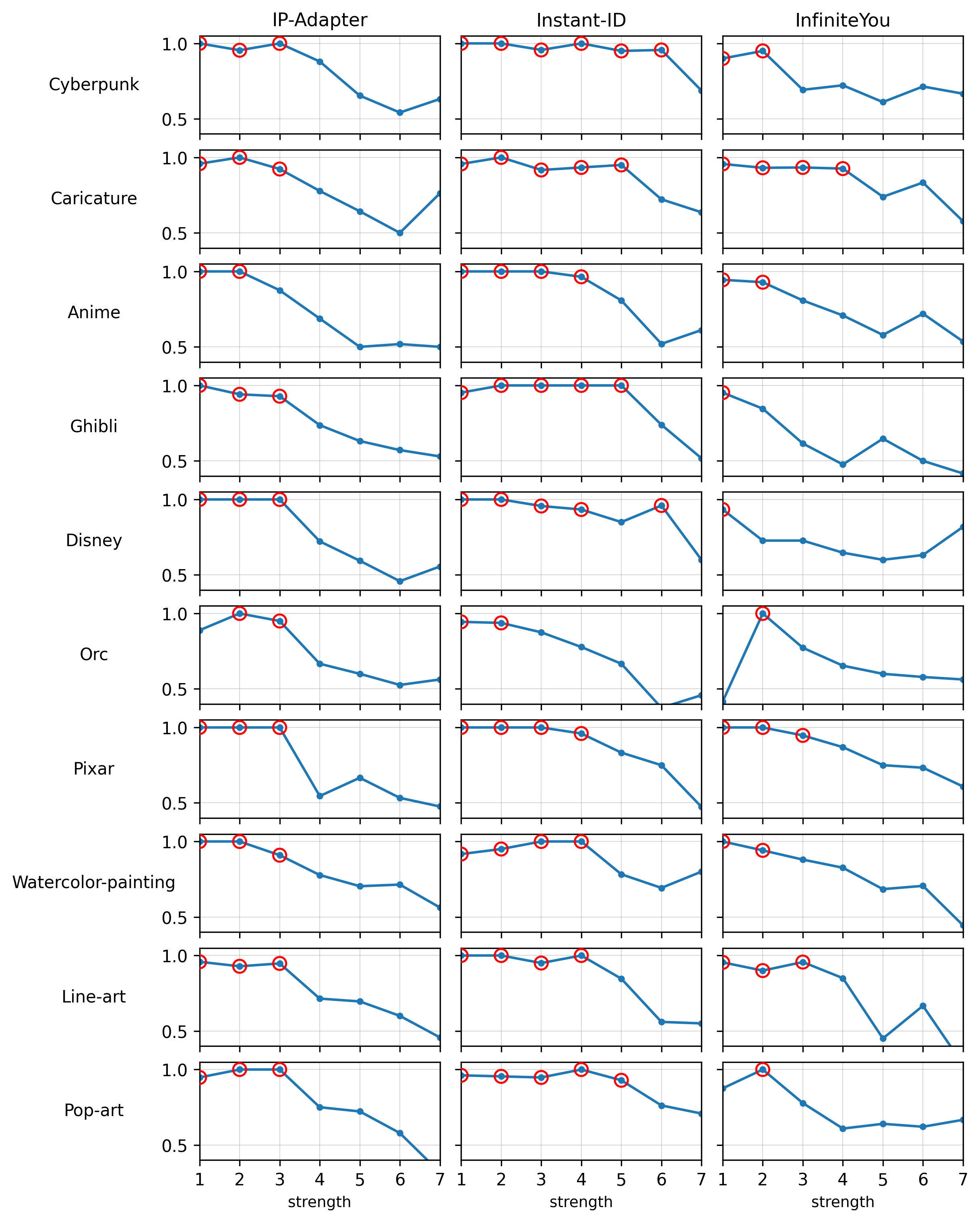} \vspace{-2mm}
  \caption{Full recognition accuracy against stylization strength for StyleBench-S. The stylization strength and style above threshold to generate StyleBench-S are circled in red. }
  \label{fig:id-s-full}
\end{figure*}

\textbf{Full recognition statistics.} We report the complete recognition accuracy as a function of stylization strength in Fig.~\ref{fig:id-s-full}. These curves visualize how different stylization methods and styles vary in perceptual difficulty and motivate our threshold-based selection strategy for constructing StyleBench-S.

\subsection{Reproducibility Notes}
\label{sec:appendix_repro}
All datasets are constructed with fixed style sets, discrete strength levels, and consistent pair formatting. The questionnaires (Figs.~\ref{fig:us-1} and~\ref{fig:us-2}) specify the instructions, ensuring that the human studies can be reproduced or extended with additional styles and methods under the same protocol. Also, all our \fixq{datasets} of StyleBench-H and StyleBench-S, and pretrained \fixq{StyleID} model will be \fixq{publicly} available upon acceptance.

\section{Additional Experiments}
\label{sec:appendix_exp}

\subsection{Receiver Operating Characteristic Curve}
We report ROC curves on both StyleBench-H and SKSF-A in Figure~\ref{fig:roc_curve}. Although SKSF-A achieves higher AUROC values, indicating stronger overall separability, StyleBench-H exhibits consistently higher true positive rates in the low–false positive rate regime, which is more relevant for high-precision identity verification. To better illustrate this behavior, we provide both linear-scale and log-scale plots, where the former highlights the overall AUROC trend, while the latter emphasizes differences under strict operating thresholds.

\subsection{Additional Comparison}

In the main paper, we report baseline comparisons using a fine-tunable ArcFace checkpoint provided by the original authors, which allows gradient-based adaptation. However, ArcFace has multiple widely used variants, including AntelopeV2, which is commonly adopted in practical systems but is distributed in a non-trainable format. To provide a more comprehensive comparison, we conduct additional experiments using the ArcFace AntelopeV2 model as a frozen encoder. We evaluate its performance under the same experimental protocols as in the main paper, enabling a fair assessment of how a widely deployed, off-the-shelf face recognition model behaves under stylization. This analysis further highlights the limitations of conventional identity encoders under strong appearance shifts and reinforces the need for a style-robust, human-aligned alternative such as StyleID.

As shown in Table~\ref{tab:compare_additional}, StyleID outperformed ArcFace AntelopeV2 across all evaluation metrics, including TPR, accuracy, and AUROC, on both StyleBench-H and SKSF-A. Notably, AntelopeV2 exhibited substantial degradation under stylization, despite its strong performance on natural photographs, whereas StyleID maintained consistently high performance across both datasets. These results further demonstrate that strong performance on photo-domain face recognition benchmarks does not directly translate to robustness under appearance shifts, and they underscore the importance of explicitly calibrating identity encoders for stylized and cross-domain scenarios.

\begin{table}[t]
\centering
\setlength{\tabcolsep}{3pt}
\caption{Results of backbone replacement experiments on \textbf{StyleBench-H} and \textbf{SKSF-A}. Each model is fine-tuned on StyleBench-S. Best results are denoted in \textbf{bold}.}
\begin{tabular}{lcccccc}
\toprule
& \multicolumn{3}{c}{\textbf{StyleBench-H}} 
& \multicolumn{3}{c}{\textbf{SKSF-A}} \\
\cmidrule(lr){2-4} \cmidrule(lr){5-7}
Methods 
& TPR$\uparrow$ & Acc$\uparrow$ & AUROC$\uparrow$
& TPR$\uparrow$ & Acc$\uparrow$ & AUROC$\uparrow$ \\
\midrule
\fixq{AntelopeV2}  
& 0.7756 & 0.7871 & 0.9566 & 0.7633 & 0.5597 & 0.9704 \\ 
StyleID 
& \textbf{0.9020} & \textbf{0.9054} & \textbf{0.9711}
& \textbf{0.8891} & \textbf{0.7393}  & \textbf{0.9922} \\ 
\bottomrule
\end{tabular}
\label{tab:compare_additional}
\end{table}

\subsection{Additional Metric Results}
To provide a more rigorous and comprehensive evaluation, we additionally report TPR at multiple FPR operating points. While the main paper reports TPR at FPR = $10^{-2}$, a setting already reflecting the substantial difficulty of identity verification under stylization, we further evaluate performance at stricter operating points of FPR = $10^{-3}$ and $10^{-4}$ under the same experimental protocol on StyleBench-H. As shown in Table~\ref{tab:additional_metrics}, StyleID consistently achieved the strongest performance across all operating points. ArcFace and AdaFace followed with moderate performance, whereas CLIP and SigLIP performed poorly under these stringent conditions. These results further confirm the robustness of StyleID under severe appearance shifts and highlight the limitations of conventional identity encoders in stylized domains.

\begin{figure}[t]
  \includegraphics[width=0.9\linewidth]{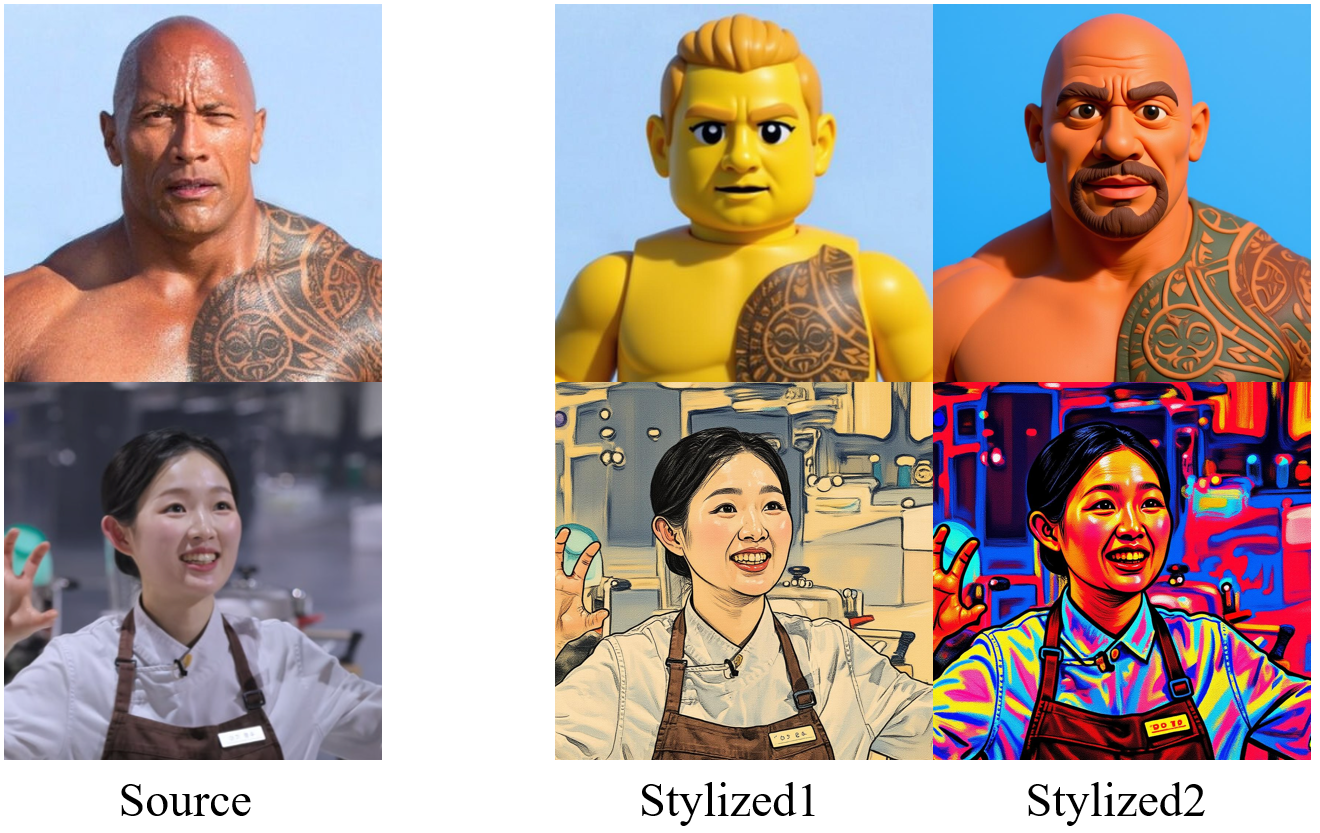}
  \vspace{-1mm}
  \caption{Examples from the user study. Participants were asked to choose which of the two stylized images better preserves the identity of the source image.}
  \label{fig:usrsty}
\end{figure}

\subsection{User Study}
We conducted a pilot user study using Flux.1 Kontext~\cite{batifol2025flux}, a flow-based image editing network, to evaluate whether the identity similarity predicted by StyleID aligns with human judgments under unseen and challenging stylization conditions. Because Flux.1 Kontext was used to generate novel styles unobserved in standard training or evaluation benchmarks, this study serves as a strict test of generalization. For each source image, we generated two stylized variants with distinct artistic styles. We recruited 15 participants (8 male, 7 female; mean age = 27.1 years) and asked them to select the image that better preserved the original identity, as illustrated in Figure~\ref{fig:usrsty}. The study consisted of 20 distinct trials. Comparing the model’s predictions with human preferences yielded an accuracy of 0.707, Cohen’s $\kappa$ of 0.392, and MCC of 0.402~\cite{landis1977measurement,matthews1975comparison}. These results indicate that the model captures meaningful identity cues even under previously unseen and difficult stylization settings, supporting its robustness and practical applicability in real-world creative scenarios.



\begin{table}[t]
\centering
\setlength{\tabcolsep}{6pt}
\caption{Results of Baseline comparison with additional metrics. Best results are denoted in \textbf{bold}.}
\begin{tabular}{lccc}
\toprule
Methods 
& TPR@10$^{-2}$ $\uparrow$  & TPR@10$^{-3}$ $\uparrow$ & TPR@10$^{-4}$$\uparrow$ \\
\midrule
ArcFace  
& 0.7649 & 0.7052 & 0.6533 \\
AdaFace  
& 0.7835 & 0.7181 & 0.7181 \\
CLIP
& 0.2560 & 0.0998 & 0.082 \\
SigLIP2
& 0.1736 & 0.0614 & 0.036 \\
StyleID 
& \textbf{0.9020} & \textbf{0.8484} & \textbf{0.8320} \\
\bottomrule
\end{tabular}
\label{tab:additional_metrics}
\end{table}

\clearpage

\end{document}